\documentclass[journal]{IEEEtran}

\ifCLASSINFOpdf
\else
\fi

\usepackage{amsmath}
\usepackage{array}
\ifCLASSOPTIONcompsoc
  \usepackage[caption=false,font=normalsize,labelfont=sf,textfont=sf]{subfig}
\else
  \usepackage[caption=false,font=footnotesize]{subfig}
\fi

\usepackage{fixltx2e}
\usepackage[colorlinks,urlcolor=blue,linkcolor=blue,citecolor=blue]{hyperref}
\usepackage{algorithm}
\usepackage{algorithmicx}
\usepackage{algpseudocode}
\usepackage{cite}
\usepackage{amsmath,amssymb,amsfonts}
\usepackage{graphicx}
\usepackage{textcomp}
\usepackage{booktabs}
\usepackage{multirow}
\usepackage[binary-units=true]{siunitx}
\usepackage{threeparttable}
\usepackage{float}
\usepackage{lipsum}
\usepackage{enumitem}
\usepackage{varioref}
\usepackage[export]{adjustbox}
\usepackage{perpage}
\usepackage{tabularx}
    \newcolumntype{L}{>{\raggedright\arraybackslash}X}

\usepackage[table]{xcolor}

\usepackage[normalem]{ulem}
\usepackage{epsfig}
\usepackage{multicol}
\usepackage{dblfloatfix}    
\usepackage{array}
\usepackage{makecell}
\usepackage{caption}
\usepackage[cmintegrals]{newtxmath}
\usepackage{subfig}
\usepackage{blindtext, graphicx}

\AtBeginDocument{%
  \providecommand\BibTeX{{%
    \normalfont B\kern-0.5em{\scshape i\kern-0.25em b}\kern-0.8em\TeX}}}

\renewcommand{\baselinestretch}{0.94}\normalsize

\usepackage{xcolor}

\usepackage[all]{background}
\usepackage{lipsum}
\SetBgContents{\textcolor{blue}{This is an extended version of the paper accepted to be published in IEEE (TCAD) 2020 which includes Appendices A and B}}
\SetBgPosition{-0.8cm,-9.0cm}
\SetBgOpacity{1.0}
\SetBgAngle{90.0}
\SetBgScale{1.3}

\begin{document}

\title{Power-Aware Run-Time Scheduler for Mixed-Criticality Systems on Multi-Core Platform}

\author{Behnaz~Ranjbar, 
        Tuan~D.~A.~Nguyen, 
        Alireza~Ejlali, 
        and~Akash~Kumar,~\IEEEmembership{Senior Member,~IEEE}

\thanks{Manuscript received Mar. 11, 2020; revised Jun. 11, \& Aug. 22, 2020; accepted Oct. 17, 2020. This manuscript was recommended for publication by Associate Editor Chia-Lin Yang. (\textit{Corresponding author: Akash Kumar)}}
\thanks{This work is supported in part by the German Research Foundation (DFG) within the Cluster of Excellence Center for Advancing Electronics Dresden (CFAED) at the Technische Universität Dresden.}

\thanks{B. Ranjbar and A. Kumar are with Technische Universität Dresden, 01062 Dresden, Germany. e-mail: \{behnaz.ranjbar, akash.kumar\}@tu-dresden.de}
\thanks{T.D.A. Nguyen is now with Xilinx Research Labs. The work was done when he was with the Department of Processor Design, Technische Universität Dresden, 01062 Dresden, Germany. e-mail:~duyatuan@acm.org}
\thanks{A. Ejlali is with the Department
of Computer Engineering, Sharif University of Technology, Tehran,
Iran. e-mail: ejlali@sharif.edu}
\thanks{B. Ranjbar is also with the Department
of Computer Engineering, Sharif University of Technology, Tehran,
Iran. e-mail: branjbar@ce.sharif.edu}
}

%


\maketitle

\begin{abstract}
In modern multi-core Mixed-Criticality (MC) systems, a rise in peak power consumption due to parallel execution of tasks with maximum frequency, specially in the overload situation, may lead to thermal issues, which may affect the reliability and timeliness of MC systems.
Therefore, managing peak power consumption has become imperative in multi-core MC systems. 
In this regard, we propose an online peak power and thermal management heuristic for multi-core MC systems. This heuristic reduces the peak power consumption of the system as much as possible during runtime by exploiting dynamic slack and per-cluster Dynamic Voltage and Frequency Scaling (DVFS). Specifically, our approach examines multiple tasks ahead to determine the most appropriate one for slack assignment, that has the most impact on the system peak power and temperature.
However, changing the frequency and selecting a proper task for slack assignment and a proper core for task re-mapping at runtime can be time-consuming and may cause deadline violation which is not admissible for high-criticality tasks. Therefore, we analyze and then optimize our run-time scheduler and evaluate it for various platforms. The proposed approach is experimentally validated on the ODROID-XU3 (DVFS-enabled heterogeneous multi-core platform) with various embedded real-time benchmarks. Results show that our heuristic achieves up to 5.25\% reduction in system peak power and 20.33\% reduction in maximum temperature compared to an existing method while meeting deadline constraints in different criticality modes.
\end{abstract}


\IEEEpeerreviewmaketitle

\section{Introduction}

\begin{figure*}[t]
\centering
\subfloat[A Task Graph]{\includegraphics[width=0.46\columnwidth]{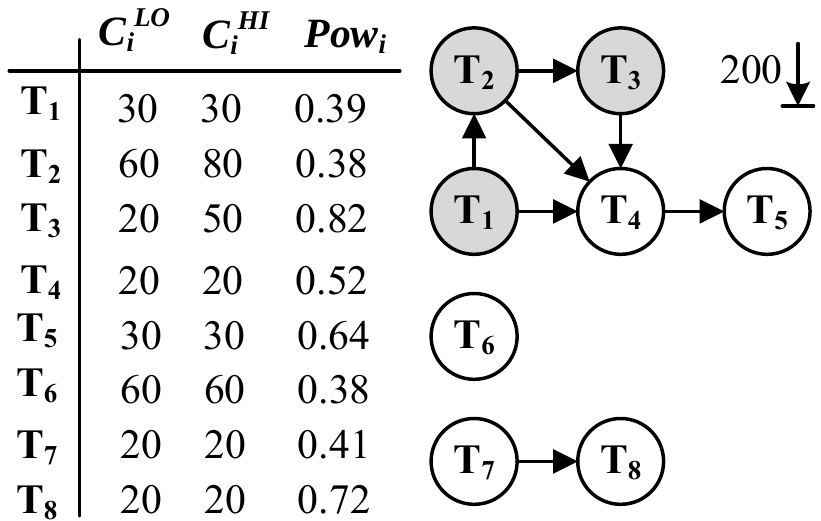}\label{fig:a_taskGraph}}
\hfil
\subfloat[system power trace at runtime without using DVFS]{\includegraphics[width=0.51\columnwidth]{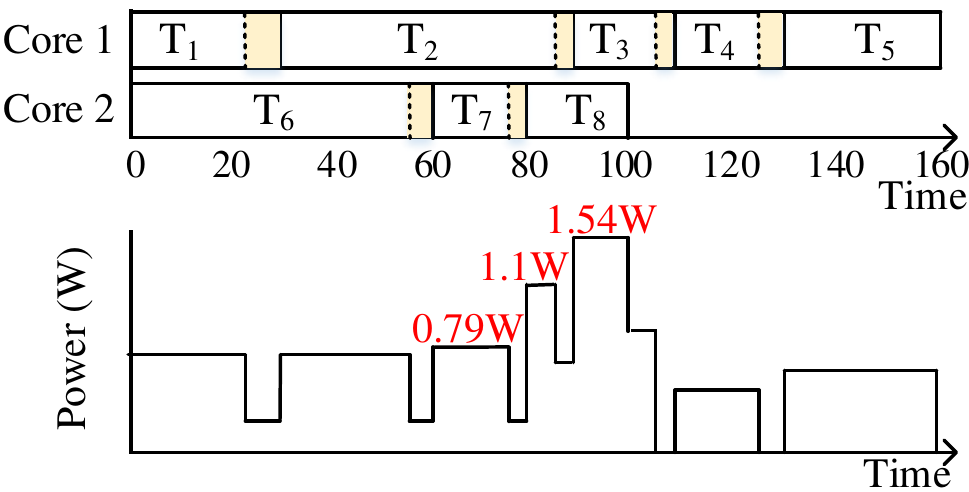}\label{fig:noDVFS}}
\hfil
\subfloat[system power trace by using DVFS and considering one task look-ahead]{\includegraphics[width=0.51\columnwidth]{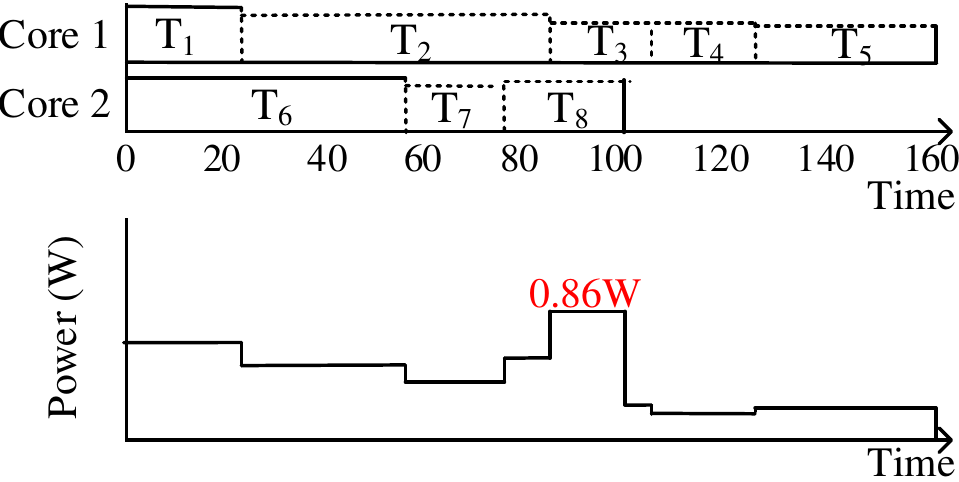}\label{fig:DVFS1}}
\hfil
\subfloat[system power trace by using DVFS and considering two tasks look-ahead]{\includegraphics[width=0.51\columnwidth]{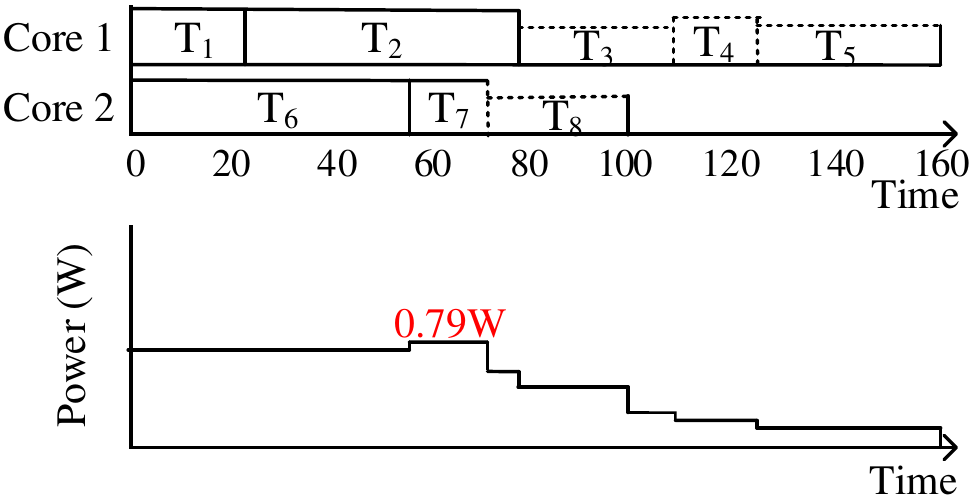}\label{fig:DVFS2}}

\caption{A motivational example for a \textbf{real-life application} in different scenarios.}
\label{fig:MExamp}
\vspace{-8pt}
\end{figure*}

\IEEEPARstart{M}{ixed-Criticality} (MC) systems are getting more attention in the last decade due to its significance in safety-critical applications (medical, flight control, etc.). In these applications, tasks are classified into multiple criticality levels in order to maintain the predictability of the applications under different unexpected behaviors~\cite{Baruah2012a,Su2013,Burns2017}. The criticality of the tasks is based on their importance and functionality to the application. For instance, the Unmanned Air Vehicle (UAV) controller is an example of an MC system~\cite{Medina2018} shown in Fig.~\ref{fig:a_taskGraph}, in which tasks have different criticality levels. In this application, the tasks with higher criticality (HC) (shown by gray color in Fig.~\ref{fig:a_taskGraph}) are responsible for the collision avoidance, navigation, and stability of the system. Failure in the execution of these tasks may lead to system failure, and cause irreparable damage to the system. The roles of low-criticality tasks (LC tasks) (shown by white color) are recording sensors data, GPS coordination, and video transmissions, which help the system to carry out its mission successfully. 
In these MC systems, to guarantee the safety of systems, HC tasks are analyzed with different assumptions, pessimistic and optimistic, in order to obtain different Worst-Case Execution Times (WCETs) for them \cite{Ernst2016,Ranjbar2019}. If the execution time of at least one HC task exceeds its optimistic WCET, the system switches from low-criticality (LO) to high-criticality (HI) mode. Then, all HC tasks continue their execution by considering the pessimistic WCET to guarantee the safety of the system~\cite{Baruah2016,Medina2018,Burns2017}.

In modern sophisticated MC systems, there are a large number of tasks. Therefore, multi-core platforms are utilized to cope with the high demands in performance~\cite{Awan2016}. However, these platforms require higher power to operate especially when the system switches to the HI mode. If the task scheduler is not aware of the power consumption, all cores might be activated at the same time with the highest performance. Therefore, the system will draw a significantly larger power than it is designed for. Systems with high peak power are more likely to generate unexpected heat that is beyond the cooling capacity. They will be more susceptible to failures and instability~\cite{Munawar2014}, which is not acceptable for the HC tasks and it may cause catastrophic consequences. In addition, as the degree of freedom (in terms of the availability of the cores) increases, it is not trivial to guarantee the real-time constraints while managing the system peak power. Therefore, managing the peak power consumption and maximum temperature of the multi-core system, while the deadlines of tasks are guaranteed at runtime, is crucial to be studied. In this work, we target a run-time scheduler to address this problem in the real world.

\textit{Motivational Example}: To clarify the problem and provide some insight into how a run-time scheduler can manage the peak power consumption, a motivational example is given. 
Fig.~\ref{fig:a_taskGraph} shows a precedence constraint MC task graph with eight tasks mapped on two cores, and tasks' information such as WCETs and peak power consumption. Although, each task in the graph can have a local deadline, the whole task graph has the deadline of \textit{D}~=~200~\textit{ms}. In addition, in order to simulate the variation in the actual run-time, these values are selected from a uniform distribution of [$\frac{2}{3}.WCET, WCET$].
We obtain the task mapping and scheduling table using the algorithm presented in~\cite{Medina2018}. In this example, we suppose that the system is only in the LO mode for simplicity of presentation. Besides, we assume that the tasks consume their maximum power continuously during their executions. Fig.~\ref{fig:noDVFS} shows the task schedule and the system power trace at runtime. In the worst-case scenario, the system's peak power may be high and may lead to thermal hotspots and instability, which has not been investigated in recent studies in MC systems that have different criticality modes. 
As shown in Fig.~\ref{fig:noDVFS}, since the tasks may finish earlier than their WCET, for example, $T1$ at 22$ms$ while its WCET= 30$ms$, the incurred slack can be exploited and assigned to the following tasks to reduce the peak power consumption. 
In Fig.~\ref{fig:DVFS1}, these dynamic slacks are used for the immediately ready tasks (one task look ahead) to decrease the speed of its corresponding core. Some power reduction can be observed. However, in some cases, the immediate task that follows may consume much less power than the other tasks after that. Therefore, it is better to reserve that slack to the task after that if it is possible. As shown in Fig.~\ref{fig:DVFS2}, if we select the task by looking two tasks ahead, more peak power reduction can be achieved as compared to Fig.~\ref{fig:DVFS1}. Therefore, by comparing the maximum power consumption of two scenarios with using the Dynamic Voltage and Frequency Scaling (DVFS) technique by looking two tasks ahead (Fig.~\ref{fig:DVFS2}) and without using DVFS (Fig.~\ref{fig:noDVFS}), we have 48.7\% reduction in peak power consumption. In addition, we have 20.12\% and 7.94\% reduction in energy consumption and peak temperature, respectively.

\textit{Proposed Method:} In this paper, we propose a heuristic to manage peak power consumption in MC systems during runtime. To achieve this, we exploit dynamic slacks, the slack between tasks' actual completion time and their WCET, along with DVFS. There are two phases in our approach: 1) at design-time, the tasks are scheduled on each core based on the Earliest Deadline First~(EDF) algorithm, and the resulting schedule is stored to be used as a static scheduling table. This is performed for both LO and HI modes. 
In this case, the number of LC tasks that have to be dropped in the HI mode is minimized, in order to improve the overall quality of service (QoS) of the system. 2) at runtime, we examine multiple tasks in the future (look-ahead) to select the most appropriate one to assign the currently available dynamic slack. The selection is based on the impact of the tasks on the peak power and temperature of the system which is quantified by a weighted multi-objective cost function. Therefore, the speed of the core that runs the task can be decreased accordingly using per-cluster DVFS. Additionally, besides exploiting the dynamic slacks, we propose a task re-mapping technique at runtime to further improve the system temperature profile. 
However, the online scheduler's timing overhead to select an appropriate task and check the re-mapping technique to select an appropriate core are crucial for the MC systems and may cause deadline violations. Furthermore, the timing overhead of changing \textit{V-f} levels in the use of the DVFS technique is critical in run-time task scheduling. Therefore, we analyze and evaluate the effect of these overheads on the schedule of MC tasks in real multi-core platforms. We study that these overheads cannot be neglected due to their impact on meeting MC tasks' deadlines. Besides, we optimize the run-time scheduler to minimize the timing overhead.

\textit{Contributions}: In summary, the main contributions of this paper are:
\begin{itemize}

    \item An online peak power and maximum temperature management of MC systems in heterogeneous multi-core platforms while respecting deadline requirements of tasks in both LO and HI modes.
    \item A multi-task look-ahead approach to make sure that dynamic slacks are assigned to the tasks that lead to more peak power and maximum temperature reduction. 
    \item An online task re-mapping technique that exploits dynamic slacks to re-map the tasks to other cores within a cluster in order to lower the system temperature.
    \item Studying the online scheduler and DVFS governor in terms of timing overhead to provide the deadline guarantee of MC tasks during run-time phase.
    \item By measuring on a real platform, we observe that while the latency of the scheduler is minimal (less than 10 $\mu$s on average), the latency of the DVFS switching is 5.313ms on average, and thus, cannot be neglected.
\end{itemize}

\textit{Evaluation}: We evaluated our run-time scheduler on ODROID~XU3/XU4, in which there are four ARM Cortex A7 and four ARM Cortex A15. 
Experiments show that our method provides peak power reduction, peak temperature reduction and average energy saving up to 5.25\%, 20.33\% ($16.8^\circ$C) and 22.44\%, respectively, compared to recent studies in the worst-case scenario.

\textit{Organization}: The rest of the paper is organized as follows. In Section~\ref{lemma:Related}, we review related works in detail. In Section~\ref{lemma:model}, we introduce the models. The problem and our proposed method in detail are presented in Section~\ref{ProbelmState} and Section~\ref{PPMethod}, respectively. In Section~\ref{OVERHEAD}, the analysis and optimization of the run-time scheduler are studied. Finally, we analyze and conclude experiments in Sections \ref{lemma:Results} and \ref{Conclusion}, respectively.

\section{Related Works}
\label{lemma:Related}

\begin{table*}[ht]
    \begin{threeparttable}[t]
	\caption{Summary of state-of-the-art approaches}
	\label{Table: related}
	\centering
	\begin{tabular}{cccccccccccc}
	\hline 
	
		\toprule
		\multicolumn{1}{p{0.08cm}}{\centering \#} &
		\multicolumn{1}{p{2.76cm}}{\centering } &
		\multicolumn{1}{p{1.2cm}}{\centering S$/$M-Core} &
		\multicolumn{1}{p{0.5cm}}{\centering Peak Power} &
		\multicolumn{1}{p{0.5cm}}{\centering Avg. Power} &
		\multicolumn{1}{p{0.8cm}}{\centering Temp.} &
		\multicolumn{1}{p{1.7cm}}{\centering DVFS (online$/$offline)} &
		\multicolumn{1}{p{0.5cm}}{\centering MC Tasks} &
		\multicolumn{1}{p{0.5cm}}{\centering DAG Model} &
		\multicolumn{1}{p{1.62cm}}{\centering DVFS Timing Overhead}  &
		\multicolumn{1}{p{2cm}}{\centering Scheduler Timing Overhead}  &
		\multicolumn{1}{p{1cm}}{\centering Real Platform}\\

		\hline
		\hline
		\multicolumn{1}{p{0.08cm}}{\centering 1} &
		\multicolumn{1}{p{2.76cm}}{\centering Li'14,~Huang'14,~Li'16, Taherin'18\cite{Guo2014,Huang2014,li2016,Taherin2018}} &
		\multicolumn{1}{p{1.2cm}}{\centering S-Core} &
		\multicolumn{1}{p{0.5cm}}{\centering \textcolor{red}{$\times$}} &
		\multicolumn{1}{p{0.5cm}}{\centering \textcolor{blue}{$\checkmark$}} &
		\multicolumn{1}{p{0.8cm}}{\centering \textcolor{red}{$\times$}} &
		\multicolumn{1}{p{1.7cm}}{\centering offline} &
		\multicolumn{1}{p{0.5cm}}{\centering \textcolor{blue}{$\checkmark$}} &
		\multicolumn{1}{p{0.5cm}}{\centering \textcolor{red}{$\times$}} &
		\multicolumn{1}{p{1.62cm}}{\centering \textcolor{red}{$\times$}}  &
		\multicolumn{1}{p{2cm}}{\centering \textcolor{red}{$\times$}}   &
		\multicolumn{1}{p{1cm}}{\centering \textcolor{red}{$\times$}}  \\

        \hline
        \multicolumn{1}{p{0.08cm}}{\centering 2} &
        \multicolumn{1}{p{2.76cm}}{\centering Awan'16~\cite{Awan2016}} &
		\multicolumn{1}{p{1.2cm}}{\centering M-Core} &
		\multicolumn{1}{p{0.5cm}}{\centering \textcolor{red}{$\times$}} &
		\multicolumn{1}{p{0.5cm}}{\centering \textcolor{blue}{$\checkmark$}} &
		\multicolumn{1}{p{0.8cm}}{\centering \textcolor{red}{$\times$}} &
		\multicolumn{1}{p{1.7cm}}{\centering offline} &
		\multicolumn{1}{p{0.5cm}}{\centering \textcolor{blue}{$\checkmark$}} &
		\multicolumn{1}{p{0.5cm}}{\centering \textcolor{red}{$\times$}} &
		\multicolumn{1}{p{1.62cm}}{\centering \textcolor{red}{$\times$}}  &
		\multicolumn{1}{p{2cm}}{\centering \textcolor{red}{$\times$}}   &
		\multicolumn{1}{p{1cm}}{\centering \textcolor{red}{$\times$}}  \\

        \hline
    
		\multicolumn{1}{p{0.08cm}}{\centering \cellcolor[HTML]{E3E2E2} 3} &
		\multicolumn{1}{p{2.76cm}}{\centering \cellcolor[HTML]{E3E2E2} Li'19~\cite{Li2019a}} &
		\multicolumn{1}{p{1.2cm}}{\centering \cellcolor[HTML]{E3E2E2}S-Core} &
		\multicolumn{1}{p{0.5cm}}{\centering \cellcolor[HTML]{E3E2E2}\textcolor{red}{$\times$}} &
		\multicolumn{1}{p{0.5cm}}{\centering \cellcolor[HTML]{E3E2E2}\textcolor{blue}{$\checkmark$}} &
		\multicolumn{1}{p{0.8cm}}{\centering \cellcolor[HTML]{E3E2E2}\textcolor{blue}{$\checkmark$}} &
		\multicolumn{1}{p{1.7cm}}{\centering \cellcolor[HTML]{E3E2E2}offline} &
		\multicolumn{1}{p{0.5cm}}{\centering \cellcolor[HTML]{E3E2E2}\textcolor{blue}{$\checkmark$}} &
		\multicolumn{1}{p{0.5cm}}{\centering \cellcolor[HTML]{E3E2E2}\textcolor{red}{$\times$}} &
		\multicolumn{1}{p{1.62cm}}{\centering \cellcolor[HTML]{E3E2E2} \textcolor{red}{$\times$}}  &
		\multicolumn{1}{p{2cm}}{\centering \cellcolor[HTML]{E3E2E2}\textcolor{red}{$\times$}}  &
		\multicolumn{1}{p{1cm}}{\centering \cellcolor[HTML]{E3E2E2}\textcolor{red}{$\times$}}  \\

        \hline
        
        \multicolumn{1}{p{0.08cm}}{\centering 4} &
        \multicolumn{1}{p{2.76cm}}{\centering  Munawar'14~\cite{Munawar2014}, Lee'14~\cite{Lee2014}} &
		\multicolumn{1}{p{1.2cm}}{\centering  M-Core} &
		\multicolumn{1}{p{0.5cm}}{\centering  \textcolor{blue}{$\checkmark$}} &
		\multicolumn{1}{p{0.5cm}}{\centering  \textcolor{red}{$\times$}} &
		\multicolumn{1}{p{0.8cm}}{\centering  \textcolor{red}{$\times$}} &
		\multicolumn{1}{p{1.7cm}}{\centering  \textcolor{red}{$\times$}} &
		\multicolumn{1}{p{0.5cm}}{\centering  \textcolor{red}{$\times$}} &
		\multicolumn{1}{p{0.5cm}}{\centering  \textcolor{red}{$\times$}} &
		\multicolumn{1}{p{1.62cm}}{\centering  -}  &
		\multicolumn{1}{p{2cm}}{\centering  -}  &
		\multicolumn{1}{p{1cm}}{\centering  \textcolor{red}{$\times$}}   \\

        \hline
        
        \multicolumn{1}{p{0.08cm}}{\centering 5} &
        \multicolumn{1}{p{2.76cm}}{\centering  Lee'10~\cite{Lee2010}} &
		\multicolumn{1}{p{1.2cm}}{\centering  M-Core} &
		\multicolumn{1}{p{0.5cm}}{\centering  \textcolor{blue}{$\checkmark$}} &
		\multicolumn{1}{p{0.5cm}}{\centering  \textcolor{red}{$\times$}} &
		\multicolumn{1}{p{0.8cm}}{\centering  \textcolor{red}{$\times$}} &
		\multicolumn{1}{p{1.7cm}}{\centering  \textcolor{red}{$\times$}} &
		\multicolumn{1}{p{0.5cm}}{\centering  \textcolor{red}{$\times$}}  &
		\multicolumn{1}{p{0.5cm}}{\centering  \textcolor{blue}{$\checkmark$}} &
		\multicolumn{1}{p{1.62cm}}{\centering  -}  &
		\multicolumn{1}{p{2cm}}{\centering  \textcolor{red}{$\times$}}  &
		\multicolumn{1}{p{1cm}}{\centering  \textcolor{red}{$\times$}}    \\

        \hline
        
        \multicolumn{1}{p{0.08cm}}{\centering 6} &
        \multicolumn{1}{p{2.76cm}}{\centering  Ansari'19~\cite{Ansari2018}} &
		\multicolumn{1}{p{1.2cm}}{\centering  M-Core} &
		\multicolumn{1}{p{0.5cm}}{\centering  \textcolor{blue}{$\checkmark$}} &
		\multicolumn{1}{p{0.5cm}}{\centering  \textcolor{blue}{$\checkmark$}} &
		\multicolumn{1}{p{0.8cm}}{\centering  \textcolor{red}{$\times$}} &
		\multicolumn{1}{p{1.7cm}}{\centering  offline} &
		\multicolumn{1}{p{0.5cm}}{\centering  \textcolor{red}{$\times$}} &
		\multicolumn{1}{p{0.5cm}}{\centering  \textcolor{blue}{$\checkmark$}} &
		\multicolumn{1}{p{1.62cm}}{\centering  \textcolor{red}{$\times$}}  &
		\multicolumn{1}{p{2cm}}{\centering  -}  &
		\multicolumn{1}{p{1cm}}{\centering  \textcolor{red}{$\times$}}    \\
		
		\hline
        \multicolumn{1}{p{0.08cm}}{\centering \cellcolor[HTML]{E3E2E2} 7} &
        \multicolumn{1}{p{2.76cm}}{\centering\cellcolor[HTML]{E3E2E2} Chaturvedi'14,~Kabir'16, Bao'09~\cite{Chaturvedi2014,Kabir2016,Min2018}} &
		\multicolumn{1}{p{1.2cm}}{\centering\cellcolor[HTML]{E3E2E2} M-Core} &
		\multicolumn{1}{p{0.5cm}}{\centering\cellcolor[HTML]{E3E2E2} \textcolor{red}{$\times$}} &
		\multicolumn{1}{p{0.5cm}}{\centering\cellcolor[HTML]{E3E2E2} \textcolor{blue}{$\checkmark$}} &
		\multicolumn{1}{p{0.8cm}}{\centering\cellcolor[HTML]{E3E2E2} \textcolor{blue}{$\checkmark$}} &
		\multicolumn{1}{p{1.7cm}}{\centering\cellcolor[HTML]{E3E2E2} offline} &
		\multicolumn{1}{p{0.5cm}}{\centering\cellcolor[HTML]{E3E2E2} \textcolor{red}{$\times$}} &
		\multicolumn{1}{p{0.5cm}}{\centering\cellcolor[HTML]{E3E2E2} \textcolor{blue}{$\checkmark$}} &
		\multicolumn{1}{p{1.62cm}}{\centering\cellcolor[HTML]{E3E2E2} \textcolor{red}{$\times$}}  &
		\multicolumn{1}{p{2cm}}{\centering\cellcolor[HTML]{E3E2E2} \textcolor{red}{$\times$}}  &
		\multicolumn{1}{p{1cm}}{\centering\cellcolor[HTML]{E3E2E2} \textcolor{red}{$\times$}}    \\

		\hline
		
		\multicolumn{1}{p{0.08cm}}{\centering \cellcolor[HTML]{E3E2E2} 8} &
		\multicolumn{1}{p{2.76cm}}{\centering \cellcolor[HTML]{E3E2E2} Qiu'12~\cite{Qiu2012}} &
		\multicolumn{1}{p{1.2cm}}{\centering \cellcolor[HTML]{E3E2E2} M-Core} &
		\multicolumn{1}{p{0.5cm}}{\centering \cellcolor[HTML]{E3E2E2} \textcolor{red}{$\times$}} &
		\multicolumn{1}{p{0.5cm}}{\centering \cellcolor[HTML]{E3E2E2} \textcolor{blue}{$\checkmark$}} &
		\multicolumn{1}{p{0.8cm}}{\centering \cellcolor[HTML]{E3E2E2} \textcolor{blue}{$\checkmark$}} &
		\multicolumn{1}{p{1.7cm}}{\centering \cellcolor[HTML]{E3E2E2} offline} &
		\multicolumn{1}{p{0.5cm}}{\centering \cellcolor[HTML]{E3E2E2} \textcolor{red}{$\times$}} &
		\multicolumn{1}{p{0.5cm}}{\centering \cellcolor[HTML]{E3E2E2} \textcolor{blue}{$\checkmark$}} &
		\multicolumn{1}{p{1.62cm}}{\centering \cellcolor[HTML]{E3E2E2} \textcolor{blue}{$\checkmark$}}  &
		\multicolumn{1}{p{2cm}}{\centering \cellcolor[HTML]{E3E2E2} -}  &
		\multicolumn{1}{p{1cm}}{\centering \cellcolor[HTML]{E3E2E2} \textcolor{blue}{$\checkmark$}}    \\
		
		\hline
		
		\multicolumn{1}{p{0.08cm}}{\centering \cellcolor[HTML]{E3E2E2} 9} &
		\multicolumn{1}{p{2.76cm}}{\centering \cellcolor[HTML]{E3E2E2} Hong'13~\cite{hong2013}, Chantem'10~\cite{Chantem2011}} &
		\multicolumn{1}{p{1.2cm}}{\centering \cellcolor[HTML]{E3E2E2} M-Core} &
		\multicolumn{1}{p{0.5cm}}{\centering \cellcolor[HTML]{E3E2E2} \textcolor{red}{$\times$}} &
		\multicolumn{1}{p{0.5cm}}{\centering \cellcolor[HTML]{E3E2E2} \textcolor{blue}{$\checkmark$}} &
		\multicolumn{1}{p{0.8cm}}{\centering \cellcolor[HTML]{E3E2E2} \textcolor{blue}{$\checkmark$}} &
		\multicolumn{1}{p{1.7cm}}{\centering \cellcolor[HTML]{E3E2E2} offline} &
		\multicolumn{1}{p{0.5cm}}{\centering \cellcolor[HTML]{E3E2E2} \textcolor{red}{$\times$}} &
		\multicolumn{1}{p{0.5cm}}{\centering \cellcolor[HTML]{E3E2E2} \textcolor{blue}{$\checkmark$}} &
		\multicolumn{1}{p{1.62cm}}{\centering \cellcolor[HTML]{E3E2E2} \textcolor{red}{$\times$}}  &
		\multicolumn{1}{p{2cm}}{\centering \cellcolor[HTML]{E3E2E2} \textcolor{red}{$\times$}}  &
		\multicolumn{1}{p{1cm}}{\centering \cellcolor[HTML]{E3E2E2} \textcolor{red}{$\times$}}    \\
		
		\hline
		
		\multicolumn{1}{p{0.08cm}}{\centering \cellcolor[HTML]{E3E2E2} 10} &
		\multicolumn{1}{p{2.76cm}}{\centering \cellcolor[HTML]{E3E2E2} Chen'6,~Kang'10,~Zhu'03, Singh'13,~Zhang'16,~Martins'17~\cite{JJChe2006,kang2010,DZhu2003,AmitKumar2013,zhang2016,Martins2017} }&
		\multicolumn{1}{p{1.2cm}}{\centering \cellcolor[HTML]{E3E2E2} M-Core} &
		\multicolumn{1}{p{0.5cm}}{\centering \cellcolor[HTML]{E3E2E2} \textcolor{red}{$\times$}} &
		\multicolumn{1}{p{0.5cm}}{\centering \cellcolor[HTML]{E3E2E2} \textcolor{blue}{$\checkmark$}} &
		\multicolumn{1}{p{0.8cm}}{\centering \cellcolor[HTML]{E3E2E2} \textcolor{red}{$\times$}} &
		\multicolumn{1}{p{1.7cm}}{\centering \cellcolor[HTML]{E3E2E2} online} &
		\multicolumn{1}{p{0.5cm}}{\centering \cellcolor[HTML]{E3E2E2} \textcolor{red}{$\times$}} &
		\multicolumn{1}{p{0.5cm}}{\centering \cellcolor[HTML]{E3E2E2} \textcolor{blue}{$\checkmark$}} &
		\multicolumn{1}{p{1.62cm}}{\centering \cellcolor[HTML]{E3E2E2} \textcolor{red}{$\times$}}  &
		\multicolumn{1}{p{2cm}}{\centering \cellcolor[HTML]{E3E2E2} \textcolor{red}{$\times$}}   &
		\multicolumn{1}{p{1cm}}{\centering \cellcolor[HTML]{E3E2E2} \textcolor{red}{$\times$}}    \\
		
		\hline

		\multicolumn{1}{p{0.08cm}}{\centering 11} &
		\multicolumn{1}{p{2.76cm}}{\centering  Chisholm'17,~Sigrist'15, Herman'12~\cite{Chisholm2017,Sigrist2015,Herman2012}} &
		\multicolumn{1}{p{1.2cm}}{\centering  M-Core} &
		\multicolumn{1}{p{0.5cm}}{\centering  \textcolor{red}{$\times$}} &
		\multicolumn{1}{p{0.5cm}}{\centering  \textcolor{red}{$\times$}} &
		\multicolumn{1}{p{0.8cm}}{\centering  \textcolor{red}{$\times$}} &
		\multicolumn{1}{p{1.7cm}}{\centering  \textcolor{red}{$\times$}} &
		\multicolumn{1}{p{0.5cm}}{\centering  \textcolor{blue}{$\checkmark$}} &
		\multicolumn{1}{p{0.5cm}}{\centering  \textcolor{red}{$\times$}} &
		\multicolumn{1}{p{1.62cm}}{\centering  -}  &
		\multicolumn{1}{p{2cm}}{\centering  \textcolor{blue}{$\checkmark$}}   &
		\multicolumn{1}{p{1cm}}{\centering  \textcolor{red}{$\times$}}    \\
		
		\hline
		
		\multicolumn{1}{p{0.08cm}}{\centering 12} &
		\multicolumn{1}{p{2.76cm}}{\centering  Trub'17~\cite{Trub2017}} &
		\multicolumn{1}{p{1.2cm}}{\centering  M-Core} &
		\multicolumn{1}{p{0.5cm}}{\centering  \textcolor{red}{$\times$}} &
		\multicolumn{1}{p{0.5cm}}{\centering  \textcolor{red}{$\times$}} &
		\multicolumn{1}{p{0.8cm}}{\centering  \textcolor{red}{$\times$}} &
		\multicolumn{1}{p{1.7cm}}{\centering  \textcolor{red}{$\times$}} &
		\multicolumn{1}{p{0.5cm}}{\centering  \textcolor{blue}{$\checkmark$}} &
		\multicolumn{1}{p{0.5cm}}{\centering  \textcolor{red}{$\times$}} &
		\multicolumn{1}{p{1.62cm}}{\centering  -}  &
		\multicolumn{1}{p{2cm}}{\centering  \textcolor{blue}{$\checkmark$}}   &
		\multicolumn{1}{p{1cm}}{\centering  \textcolor{blue}{$\checkmark$}}    \\
		
		\hline
		\multicolumn{1}{p{0.08cm}}{\centering \cellcolor[HTML]{E3E2E2} 13} &
		\multicolumn{1}{p{2.76cm}}{\centering \cellcolor[HTML]{E3E2E2} Guo'19~\cite{Guo2019}} &
		\multicolumn{1}{p{1.2cm}}{\centering \cellcolor[HTML]{E3E2E2} M-Core} &
		\multicolumn{1}{p{0.5cm}}{\centering \cellcolor[HTML]{E3E2E2} \textcolor{red}{$\times$}} &
		\multicolumn{1}{p{0.5cm}}{\centering \cellcolor[HTML]{E3E2E2} \textcolor{blue}{$\checkmark$}} &
		\multicolumn{1}{p{0.8cm}}{\centering \cellcolor[HTML]{E3E2E2} \textcolor{red}{$\times$}} &
		\multicolumn{1}{p{1.7cm}}{\centering \cellcolor[HTML]{E3E2E2} online} &
		\multicolumn{1}{p{0.5cm}}{\centering \cellcolor[HTML]{E3E2E2} \textcolor{red}{$\times$}} &
		\multicolumn{1}{p{0.5cm}}{\centering \cellcolor[HTML]{E3E2E2} \textcolor{blue}{$\checkmark$}} &
		\multicolumn{1}{p{1.62cm}}{\centering \cellcolor[HTML]{E3E2E2} \textcolor{blue}{$\checkmark$}}  &
		\multicolumn{1}{p{2cm}}{\centering \cellcolor[HTML]{E3E2E2} \textcolor{blue}{$\checkmark$}}   &
		\multicolumn{1}{p{1cm}}{\centering \cellcolor[HTML]{E3E2E2} \textcolor{blue}{$\checkmark$}}    \\

		\hline
		
		\multicolumn{1}{p{0.08cm}}{\centering 14} &
		\multicolumn{1}{p{2.76cm}}{\centering   Our Work} &
		\multicolumn{1}{p{1.2cm}}{\centering  M-Core} &
		\multicolumn{1}{p{0.5cm}}{\centering  \textcolor{blue}{$\checkmark$}} &
		\multicolumn{1}{p{0.5cm}}{\centering  \textcolor{blue}{$\checkmark$}} &
		\multicolumn{1}{p{0.8cm}}{\centering  \textcolor{blue}{$\checkmark$}} &
		\multicolumn{1}{p{1.7cm}}{\centering  online}   &
		\multicolumn{1}{p{0.5cm}}{\centering  \textcolor{blue}{$\checkmark$}} &
		\multicolumn{1}{p{0.5cm}}{\centering  \textcolor{blue}{$\checkmark$}} &
		\multicolumn{1}{p{1.62cm}}{\centering  \textcolor{blue}{$\checkmark$}}  &
		\multicolumn{1}{p{2cm}}{\centering  \textcolor{blue}{$\checkmark$}}   &
		\multicolumn{1}{p{1cm}}{\centering  \textcolor{blue}{$\checkmark$}} \\
		
		\bottomrule
	\end{tabular}
	\end{threeparttable}
\end{table*}

Many previous works in the context of MC systems have just focused on proposing techniques in the field of task scheduling and mapping in both online and offline phases. 
Since our focus is on online MC task scheduling to manage power and temperature, we only consider the works presented for MC or non-MC systems with similar scope. 
Generally, the related works on power and thermal management for real-time systems can be classified based on the assumed platform, single (S) or multi (M)-core, MC or non-MC systems and the target optimization objectives of peak power, average power or maximum temperature. Table~\ref{Table: related} summarizes the recent works with different target optimization objectives. 

From the perspective of power and thermal management in MC systems, some works such as~\cite{Guo2014,Huang2014,li2016,Taherin2018,Awan2016} have presented methods to minimize the average power consumption in MC systems theoretically in which systems are single or multi-core (rows 1 and 2 in Table~\ref{Table: related}). 
In general, they only optimize the average power in the LO mode in simulation. 
When the system switches to the HI mode, all HC tasks are executed with the highest frequency; and all LC tasks are dropped. As a result, in the HI mode, with higher frequency, the peak power consumption of the system may increase significantly, which is not admissible. 
Furthermore, there is a paper~\cite{Li2019a} that has considered thermal management in MC systems (third row of the Table~\ref{Table: related}). The researchers minimize the temperature of single-core processors by finding the optimum speed for each task in the design-time phase. Hence, they discard LC tasks when the system switches to the HI mode, which is not acceptable in many MC applications. Besides, they do not consider the latency of changing the \textit{V-f} level at runtime, which may cause deadline violation and, consequently, catastrophic consequences. 

As we focus on peak power management, some studies concentrate on peak power management in multi-core systems at design-time (rows 4-6). These papers have only considered hard real-time tasks with one criticality level which is not practical for MC.
It should be mentioned that authors in~\cite{Lee2010} work on the dependent task model in which the execution of some tasks is postponed to manage the simultaneous peak power consumption. It is not suitable for MC tasks, especially in the HI mode.

The previous works in the context of power or thermal management in non-MC systems that use DVFS by considering the dependent task model are shown in rows (7-10) of Table~\ref{Table: related}. Researchers in~\cite{Amit2020} give a comprehensive study in the field of energy and thermal management of multi-core platforms. 
Since our approach is reducing the power and temperature at runtime, we review the works that have used the online DVFS, which are~\cite{Min2018,kang2010,DZhu2003,AmitKumar2013,zhang2016,JJChe2006,Martins2017}. In~\cite{Min2018}, a look-up table for each task is generated in the offline phase, which contains the optimum voltage and frequency settings for each core for every possible run-time condition, task execution time, and core temperature measurement. The memory overhead incurred in generating these tables may not be desirable, especially for multi-core systems with many tasks and cores. 
Researchers in~\cite{kang2010,DZhu2003,AmitKumar2013,zhang2016,JJChe2006,Martins2017} have used slack reclamation to apply online DVFS to the system while executing dependent tasks. Kang et al.~\cite{kang2010} propose an algorithm that uses dynamic voltage scaling to minimize energy without considering the tasks' deadlines, which is not suitable for MC systems. Researchers in ~\cite{DZhu2003,AmitKumar2013,zhang2016} suggest a run-time energy management technique that uses reclaimable slack for the immediately ready task to decrease average power. Their results show that the power can be reduced; however, the possibilities of looking further ahead into the future execution of the following tasks to have better results are not explored. Besides, in~\cite{Martins2017}, the authors have considered two types of tasks, best effort and real-time, and they have just used the dynamic slack for the next real-time task to reduce its \textit{V-f} level, which is inefficient. There is an aggressive slack reclamation algorithm, presented by~\cite{JJChe2006}, in which, the generated dynamic slack is checked to be able to use for the next task if the remaining tasks could complete their execution before the deadline. However, in general, the average energy consumption is reduced, but this algorithm has focused more on meeting the deadlines, while we target both energy minimization and meeting the deadlines.

Most of the related works that we discussed here, have evaluated their method in a simulation, and just focused on power or temperature minimization, and are not concerned about run-time behaviour and its associated timing overhead. 
On the other hand, there are some previous works in the context of MC systems that just focus on considering timing overhead of scheduler~\cite{Chisholm2017,Sigrist2015,Trub2017,Herman2012}, memory sharing and bus communication latency~\cite{Giannopoulou2014,Chisholm2015,Chisholm2016,Kim2018} to evaluate their method and have a real implementation. Since our focus is on online MC task scheduling, we only consider works that have studied the scheduling latency.
Most of the papers, which have considered the latency of the scheduler in the field of MC systems (row 11-12 in Table~\ref{Table: related}), are limited to the timing overhead of task monitoring, task termination and arriving and mode switching, not the scheduler. Further, these run-time overheads are measured on the Intel platform, i.e., not embedded processors.

In addition, from the DVFS latency perspective, some few works, e.g.,~\cite{Guo2019}, have presented a method to minimize energy in a multi-core platform by using the DVFS technique. Researchers in~\cite{Guo2019} have considered a task graph model running on the cluster-based platform. They have also considered the latency of changing frequency in their paper. As shown in Table~\ref{Table: related} row 13, they have not considered peak power or thermal management and also, their method is not suitable for MC systems where tasks have different criticality levels.

We study run-time scheduling of dependent MC tasks, which are executed on a multi-core platform to manage peak power and maximum temperature by considering the timing overheads such as frequency changing and run-time scheduler, which is not found in existing MC works.

\section{System Models}
\label{lemma:model}

\subsection{Task Model}
\label{lemma:taskmodel}
We consider real-time applications consisting of dependent periodic MC tasks, such that, each task $\tau_i$ is represented as \{$\zeta_i, C_i^{LO}, C_i^{HI}, d_i, Su_i, Pr_i$\}. Analogous to~\cite{Medina2018,Baruah2016}, we consider dual-criticality system where each dependent MC task can be either high-critical ($\zeta_i$~=~HC) or low-critical ($\zeta_i$~=~LC). Further, each task $\tau_i$ has a deadline $d_i$. The successors and predecessors of each task are determined by $Su_i$ and $Pr_i$, respectively. A task can be executed after all its predecessor tasks have finished their execution.
Each MC task has different WCETs, $C_i^{LO}$ (optimistic) and $C_i^{HI}$ (pessimistic) that for each LC task $C_i^{LO}=C_i^{HI}$ and also, for each HC task $C_i^{LO} \leq C_i^{HI}$. 
If a task is a predecessor of an HC task, then it is considered as an HC task as well. In addition, all tasks have a common period $P$ which is the period of the task graph.

In general, MC systems have two modes of operation: LO and HI. Initially, the system starts in the LO mode in which all LC and HC tasks must be executed correctly before their deadlines. When the execution time of at least one HC task exceeds its $C_i^{LO}$ due to unexpected conditions, the system switches to the HI mode and then all HC tasks are executed with their $C_i^{HI}$. In the dependent MC task model, the system switches back safely to the LO mode at the end of each period~\cite{Medina2018, Baruah2016}.

\subsection{Hardware Architecture Model}
\label{SystemModel}

We consider a multi-core processor comprising of $m$ cores \{$C_1, C_2, ..., C_m$\} based on the ODROID~XU3 board, where the system is DVFS-enabled and the cores can operate at multiple voltage (\textit{V}) and frequency (\textit{f}) levels. The ODROID~XU3 consists of two clusters with ARM cortex-A15 (big) and ARM Cortex-A7 (LITTLE) (four big cores and four LITTLE cores); hence, cores within the same cluster operate at the same \textit{V-f} level and also, each cluster can operate at different frequency and voltage levels. In this board, the allowed frequency is in the range of [0.2,~1.4]~\textit{GHz} for LITTLE cores and [0.2,~2]~\textit{GHz} for big cores. Besides, the voltage is in the range of [0.9, 1.3]~\textit{V} for LITTLE cores and [0.9, 1.3625]~\textit{V} for big cores.

\subsection{Power Model}
\label{Powermodel}
The total power consumption of a core is composed of static ($P_{s}$), dynamic ($P_{d}$) and independent power consumption ($P_{ind}$)~\cite{Taherin2018,li2016}. $P_{ind}$ refers to the power related to the memory and I/O activities. As mentioned in Section~\ref{SystemModel}, the \textit{V-f} level of an entire cluster can be changed. This implies that all cores in a cluster must have the same \textit{V-f} level. The total power consumption is given by Eq.~\eqref{eq:1}. In this equation, $I_{sub}$ and $C_L$ are the sub-threshold leakage current and load capacitance, respectively. In this paper, we focus on decreasing $P_d$. 
\begin{equation}
\label{eq:1}
P=P_{s}+P_{d}+P_{ind}= I_{sub}V+C_{L}V^{2}f+P_{ind}
\end{equation}
in which: ($\rho_1$ and $\rho_2$ are the scaling factors of frequency and voltage, respectively) 
\begin{equation}
\label{eq:cont1}
f_{min} \leq f=\rho_1 \times f_{max} \leq f_{max} ,\\ 
V_{min} \leq V=\rho_2 \times V_{max} \leq V_{max} \nonumber
\end{equation}

Therefore, by using these scaling factors, Eq. \ref{eq:1} can be written based on the $V_{max}$ and $f_{max}$ as:
\begin{equation}
\label{eq:rhomax}
P=I_{sub}({\rho_2} V_{max})+C_{L}({{\rho_2} V_{max})}^{2} ({\rho_1} f_{max})+P_{ind}
\end{equation}

As our system is based on the ODROID~XU3, some frequency levels work with the same voltage level on this board. It means, by reducing the frequency level, the voltage level does not change. Therefore, the scaling factors $\rho_1$ and $\rho_2$ do not have the same value. According to the range of frequency for big and LITTLE cores presented in Section~\ref{SystemModel}, $\rho_1$ can be set in the range of [0.143,~1] for the A7 cores and [0.1,~1] for A15 cores. In addition, $\rho_2$ is in the range of [0.692,~1] and [0.6606,~1] for A7 and A15 cores, respectively. Although the ODROID~XU3 has power sensors, they only report values for the entire cluster, not for each core.

\section{Research Objectives}
\label{ProbelmState}
We target peak power consumption and maximum temperature issues in MC systems and evaluate the algorithm on a real multi-core platform. Although there are works that manage or minimize the power consumption of MC systems as previously discussed in Section~\ref{lemma:Related}, they have not considered the instantaneous peak power consumption in both HI and LO modes and their algorithms have often been limited to simulation. 
One of the most common approaches to solve the problem is to exploit the dynamic slack generated at runtime to change \textit{V-f} levels of cores, while the MC tasks' deadlines are guaranteed. 
The crucial research questions that are addressed in our article are as follows: (1)~How to select the most appropriate tasks to assign the dynamic slack to, for managing the peak power consumption; (2)~Whether it is possible to re-map the tasks to other cores for better thermal control, and if yes, where and when should the tasks be re-mapped to; (3)~Which timing overheads during runtime have an impact on task scheduling and deadline misses; (4)~How these run-time timing overheads can be managed to not affect tasks' deadlines.

\section{Proposed Method: Online Peak Power and Max. Temperature Management Method in MC Systems}
\label{PPMethod}

The goal of our proposed method is to minimize the peak power and the maximum temperature of individual cores during run-time.
\begin{align}
 & Minimize (P_j,(T_{max})_j)|_{(j \in Cores)}  
\end{align}

DVFS is one of the techniques that we use to manage the metrics (peak power and maximum temperature). Reducing the \textit{V-f} level of a core while executing a task, increase the execution time of the task and it may cause deadline violation. In addition, the latency of changing \textit{V-f} level or run-time scheduling may cause deadline violation. 
Eq.~\ref{eq:3n} represents that the sum of the execution time of each task \textit{i} on the core \textit{j} at the \textit{V-f} level \textit{l} and timing overheads of the run-time scheduler ($TO_{Sch.}$) and changing \textit{V-f} level ($TO_{Vf}$) must not exceed the task deadline in each criticality mode.
\begin{equation}
\label{eq:3n}
TO_{Sch.}+TO_{Vf}+ \frac{C_i}{f_{jl}} \leq d_i \to \begin{cases}
                                C_i= C_i^{LO}\quad \text{if } mode=LO\\
                                C_i= C_i^{HI} \quad \text{if } mode=HI
                             \end{cases}
\end{equation}

The proposed approach consists of design-time and run-time phases. 
It is worth noting that the proposed method takes advantages of the run-time phase to manage the peak power and temperature; hence it is not possible to use any optimization method such as ILP (Integer Linear Programming) due to its long execution time. Thus, we develop a heuristic-based method. Fig.~\ref{fig:Overal} shows the flow of our proposed approach, along with the Hardware Platform. The Hardware Platform is used in Design-Time Phase for tasks' power profiling and in Run-Time Phase for execution of task on cores. Now, we explain our approach comprising of the Design-Time and Run-Time Phases, in detail.

\subsection{Design-Time Phase}
The input to the algorithm is a precedence constrained task set and the multi-core system description, as shown in Fig.~\ref{fig:Overal}. The power required by the tasks can be obtained by running the benchmarks on a real platform, which is discussed in detail in Section~\ref{lemma:Results}. It should be noted that handling an unknown application during runtime is beyond the scope of this paper. Since we target embedded applications, normally, the designer knows the system's tasks and their parameters at design-time. Therefore, by using the parameters of MC tasks such as WCETs, two tables of static task mapping and scheduling for LO and HI modes are created as shown in Design-Time Phase of Fig.~\ref{fig:Overal}. 
EDF algorithm is used to calculate the schedule of the tasks in each of the two modes statically based on the WCETs of LC and HC tasks, using the algorithm presented in~\cite{Medina2018}. 
In the LO mode, all tasks are scheduled with equal priority; in the HI mode, HC tasks are scheduled with a higher priority. 
These static schedules in the respective modes are then used to execute all tasks at run-time. This enforces a strict ordering in the execution of the tasks and guarantees that all deadlines are met according to the design-time analysis in both modes. It should be noted that since the WCETs of HC tasks are higher in the HI mode, not all LC tasks may be schedulable in the HI mode. In order to maximize the overall QoS, the algorithm tries to drop as few LC tasks as possible when computing the HI mode table. These tables and the info associated with the tasks are used during run-time phase by our algorithm to manage the system.

\subsection{Run-Time Phase}
\label{runtimephase}
\begin{figure}[t]
\centering
\includegraphics[width=1\columnwidth]{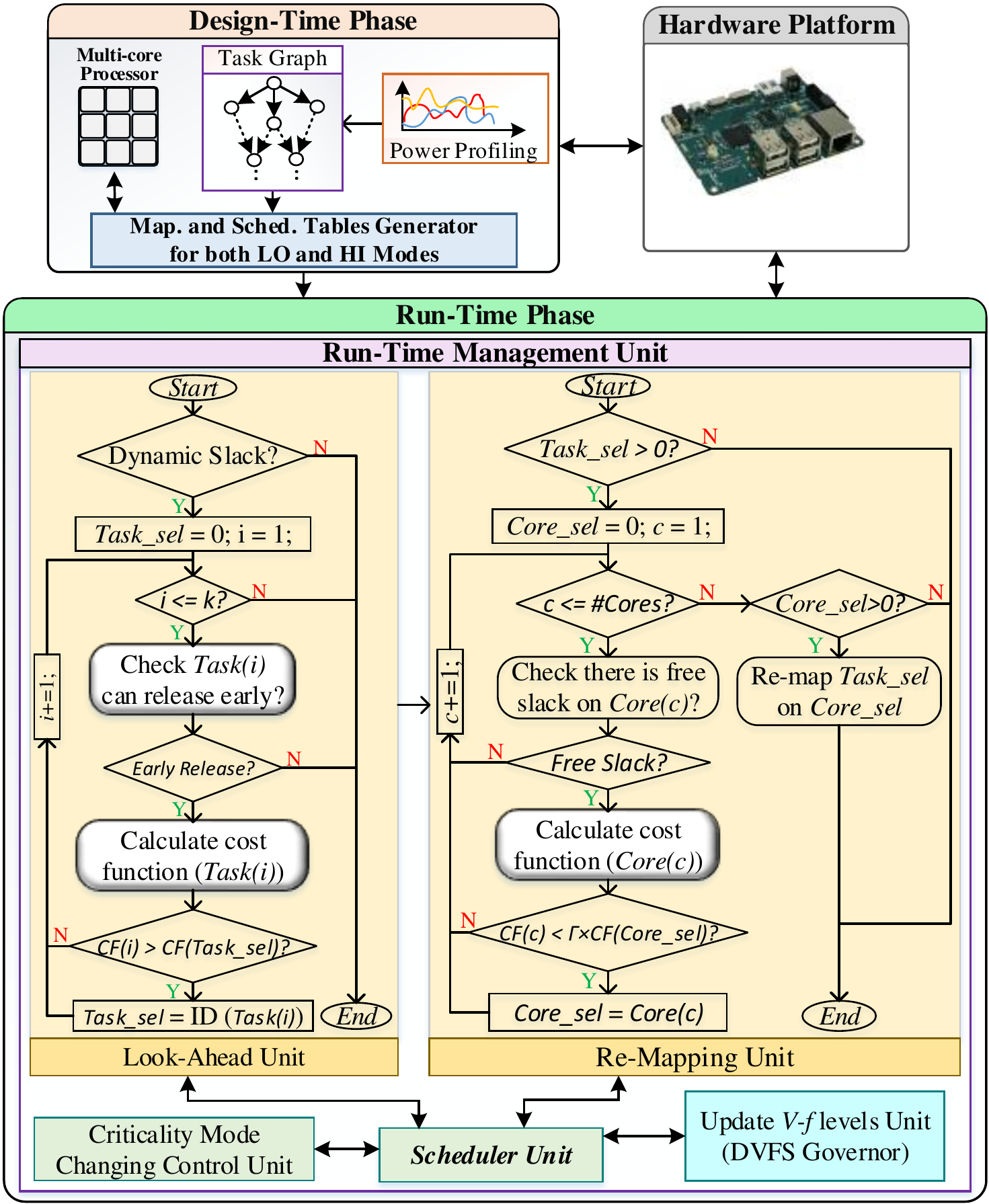}
\caption{Overview of our proposed approach.}
\label{fig:Overal}
\vspace{-7pt}
\end{figure}

The run-time phase of our proposed method consists of several function control units, as shown in Fig.~\ref{fig:Overal}. The Scheduler Unit is the main unit that is communicating with the other units. Two main functions are supported in this unit: 1) Execute the tasks according to the tables; 2) Change the scheduling and mapping of the tasks according to our proposed policy which we discuss in Sections \ref{lookaheadT} and \ref{CoreRemapT}. 
When there is any free slack on a core, or a task finishes its execution early, the Look-Ahead Unit is executed. This unit is used to choose a subset of tasks and select the most appropriate one among them. If an appropriate task is selected in a core, according to the core temperature and temperature of other cores, the Re-Mapping Unit is used to reduce the maximum temperature and decide whether to re-map the task to other cores or not. After that, the obtained \textit{V-f} level for the core is stored. This stored frequency is used by the DVFS governor Unit when the task is ready to be executed. The details of the DVFS Governor Unit to select the optimum \textit{V-f} level for a cluster is discussed in Section~\ref{governor}. Due to MC systems' behavior, the system switches to the HI mode if the execution of at least one HC task exceeds its defined $C^{LO}$. It should be checked by the Criticality Mode Changing Control Unit presented in Fig.~\ref{fig:Overal}. In this case, the system changes its task scheduling according to the HI scheduling table which is generated at design-time. The details for Look-Ahead Unit and Re-mapping Unit are described as follows.

\vspace{5pt}
\subsubsection{\textbf{Selecting the Appropriate Task to Assign Slack}}
\label{lookaheadT}
In Look-Ahead Unit, we consider an approach named look-ahead in which our algorithm chooses $k$ tasks after generated dynamic slack and also mapped on the same core in which the dynamic slack ($S_{dyn}$) is generated\footnote{Finding the optimum value for $k$ is discussed in Section~\ref{lemma:Results}.}. 
For each of the $k$ tasks, a cost function is computed as defined by Eq.~\ref{eq:4} below.
\begin{equation}
\label{eq:4}
CF_i = \alpha \times E_i + \beta \times Pow_i
\end{equation}

\begin{figure}[t]
\centering
\subfloat[before assigning slack]{\includegraphics[width=0.9\columnwidth]{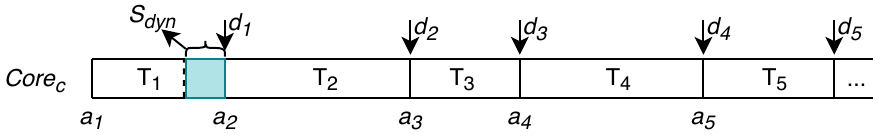}\label{fig:beforeLAU}}
\hfil
\subfloat[after assigning slack]{\includegraphics[width=0.9\columnwidth]{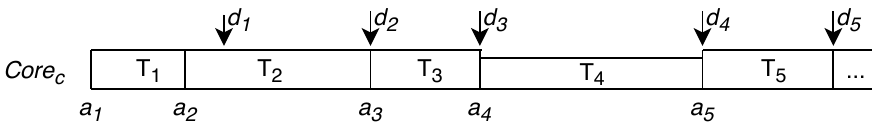}\label{fig:after}}
\caption{An example of look-ahead policy}
\label{fig:LAUExamp}
\vspace{-8pt}
\end{figure}

In this function, $Pow_i$ and $E_i$ are the maximum instantaneous power and maximum energy, respectively, that a task consumes to execute. In addition, $\alpha$ and $\beta$ are in the range of [0,1]. Besides, energy reduction leads to a decrease in chip temperature~\cite{Huang2014}. 
Note that, if we consider $\langle\alpha,~\beta\rangle$ = $\langle 0,1 \rangle$, the cost function only considers power of a task, and not it's energy. Hence, the task with the largest peak power consumption is chosen to be executed at reduced core speed, in order to reduce the peak power consumption. If we have $\langle\alpha,~\beta\rangle$~=~$\langle 1,0 \rangle$, only energy is considered in the cost function. Hence, the task with the largest energy consumption is chosen to be executed at reduced core speed, thereby reducing the maximum energy consumption 
(Appendix A shows the proof of optimal solution of peak power minimization in individual cores). 
After selecting the task, the maximum power consumption, and its WCET ($C_i^{LO}$ or $C_i^{HI}$) are changed based on the size of generated slack time and the \textit{V-f} level. As a result, the start time and the deadline of tasks that are executed between the generated dynamic slack and selected task are \textit{shifted} left based on the amount of slack to let the chosen task run with less speed, for example tasks $T_2$ and $T_3$ illustrated in Fig.~\ref{fig:LAUExamp}. 

Furthermore, Eq.~\ref{eq:4} is applied to a set of tasks that can start their executions earlier. A task ($\tau_i$) can start early if it is released before $a_i-S_{dyn}$, where $a_i$ is the start time of $\tau_i$. As mentioned, a task can be released when all its predecessors finish their execution. Therefore, we just check $T_{ri}\leq~a_i-S_{dyn}$, where $T_{ri}$ is the release time of $\tau_i$. 
Consider the selected task~$\tau_i$ with the start time~$a_i$ and deadline~$d_i$ that $a_i+WCET_i\leq~d_i$. Assuming that we have the amount of slack, $S_{dyn}$ generated by $\tau_j$, during runtime. To utilize this slack time for the appropriate task $\tau_i$, in general, the scheduler finds the minimum acceptable frequency based on $f_i=max(f_{min},\frac{WCET_i}{WCET_i+S_{dyn}}.f_{max})$. This ensures that only the start time of the task is earlier by $S_{dyn}$ and the deadline is kept unchanged, for example $T_4$ shown in Fig.~\ref{fig:LAUExamp}. Hence, $a_i-S_{dyn}+\frac{WCET_i}{f_i/f_{max}}\leq~a_i+WCET_i\leq d_i$. However, as mentioned at the beginning of this section, selecting the proper task and the core, and changing the \textit{V-f} level have overheads\footnote{We discuss in Section~\ref{lemma:Results}, how theses timing overheads ($TO_{Sch.}$ and $TO_{Vf}$) are measured.}. If we ignore them while selecting the optimum frequency, it may cause a deadline violation. Therefore, $S_{dyn}$ is reduced by $TO_{Sch.}$ and $TO_{Vf}$. After selecting the optimum frequency the start time of the appropriate task $\tau_i$ ($a_i$) is updated for the static schedule.

\vspace{5pt}
\subsubsection{\textbf{Re-Mapping Technique}}
\label{CoreRemapT}
In order to manage the maximum temperature of the system and have better thermal control, it is possible to re-map the selected task to the other cores without changing its deadline. Therefore, to decide about re-mapping the task and selecting the appropriate core to re-map, we use the cost function in Eq.~\ref{eq:5}.
\begin{equation}
\label{eq:5}
CF_c = \Gamma \times \sum_{t= 1}^{t_f}  E_c(t)
\end{equation}

In this cost function, instead of using actual core temperature, we predict their temperature according to the accumulated energy. Based on our observation, a core tends to have a lower temperature when its accumulated energy is less than the other cores\footnote{We show the observation about the relevance between core temperature and its accumulated energy consumption in Section~\ref{lemma:Results}}. However, the difference between the accumulated energy of the base core and the selected core should be large enough. Therefore, we define a coefficient ($\Gamma$), which is equal to 0.9 in our experiments. In this equation, $t_f$ is the time when any particular task is finished. 
Besides, in order to not affect the tasks' deadline mapped on other cores, cores are examined for re-mapping that have free slack at the same period to execute the appropriate task.
Since we consider the clustered multi-core platform (ODROID~XU3) for our experiment, each application's execution time and power consumption will be different when running on different clusters. Hence, we use the re-mapping technique within each cluster. 
The reason is that although re-mapping from a little core to a big core reduces a task's execution time, it causes the system peak power consumption to increase, which is not acceptable based on our targets. Therefore, to not change the system peak power consumption, we use the re-mapping technique within each cluster. 
It should be noted, since the re-mapping technique is applied to a task that is not started yet, and also, the technique is done in parallel with changing the frequency, the migration overhead does not affect the deadline constraints. The reason is that the latency of re-mapping is much less than the latency of changing the frequency, which we study in detail in Section~\ref{lemma:Results}.

\subsection{Run-Time Management Algorithm}

The pseudo-code of our proposed algorithm is outlined in Algorithm~\ref{alg:PPReduce}. At first, the algorithm gets the set of precedence constraint tasks, the number of tasks looking ahead ($k$), scheduling table for each mode, and available \textit{V-f} levels for cores as inputs. Then it gives start time and the \textit{V-f} level assignment for each task at runtime.
At the initialization step, the system starts its operation in the LO mode, and also, the voltage and frequency of each core are set to the maximum value (lines~1-3). The proposed online peak power reduction algorithm is presented in (lines~4-45). At first, the algorithm checks that whenever the execution time of a task exceeds its \textit{WCET}, the system switches to the HI mode (lines~5-9). If any task execution exceeds its $C_i^{LO}$ and the output of this task is not ready, the system switches to the HI mode and remains in this mode till the end of the period. In this situation, at the beginning, the \textit{V-f} level of each core is set to the maximum value to meet the deadline of HC tasks (lines~7-8). The rest of the algorithm is executed in both modes. 

\begin{algorithm}[t]
    \caption{Online Peak Power Reduction Algorithm}
    \label{alg:PPReduce}
    \begin{threeparttable}
		{ \small
				
			\begin{algorithmic}[1]
                \renewcommand{\algorithmicrequire}{\textbf{Input:}}
	        	\renewcommand{\algorithmicensure}{\textbf{Output:}}
					
		        \Require Task Graph ($G_{T}$), Cores, Scheduling Tables of each Mode ($Sch_L$ and $Sch_H$), Number of Tasks Looking Ahead ($k$).
					
		       \State mode $\gets$ 0 , MO $\gets$ LO; // the system starts from the LO mode  \par 
		            \hskip\algorithmicindent \hskip\algorithmicindent
		            \hskip\algorithmicindent
		            and $Sch_L$ is used to schedule the tasks
		            
		            \For {\textbf{each} core \textit{j}}	initialize the \textit{V-f} level to maximum; \
               	    \EndFor
		        
		        \Procedure{MCS Online PPReduction}{}

		                \If{each Task executes more than \textit{$C^{MO}$}}
		            
		                    \State mode $\gets$ 1, MO $\gets$ HI;  // System switches to the HI \par 
                					\hskip\algorithmicindent \hskip\algorithmicindent \hskip\algorithmicindent \hskip\algorithmicindent  mode and task scheduling is done by $Sch_H$
                		    \For {\textbf{each} core \textit{j}}	initialize the \textit{V-f} level to maximum; \
               	             \EndFor
		            
		                \EndIf

                        	\If{each Task finishes its execution earlier than its deadline \par 
                		\hskip\algorithmicindent \textbf{or} there is an idle time in a core }
                
                	\If{$Task_i$ has already finished its execution}						
		                 \State $S_{dyn}$ $\gets$ Extract\_Dynamic\_Slack(); //$C^{MO}_{i}$ - $ACT_{i}$

    \Else \textbf{if} there is an idle time in a core \textbf{then}

         \State $S_{dyn}$ $\gets$ Extract\_Dynamic\_Slack(); //idle time

	\EndIf
                            
		                    \State $T_S$, $T_P$ $\gets$ 0
		                    
		                    \State $S_{dyn}$ -= $TO_{sch}+TO_{Vf}$;
		                    
		                      \For {\textit{n} = 1 \textbf{to} $k$}

		                                \State $T_P$ $\gets$ $\tau_{n_{th}}$ after generated slack;

		                            \If{$CF_{T_S}$ $<$ $CF_{T_P}$ \textbf{and} $T_P$ can start earlier  
                					} 

		                                \State $T_S$ $\gets$ $T_P$, $n_{s}$ $\gets$ n;
		                    
		                            \EndIf
		                    
		                     \EndFor

		                        \If{$n_{s}$ $>$ 0}

		                                \State \textit{$Freq^{MO}_{T_S}$} $\gets$ max ($f_{min}$,$\frac{C^{MO}_{T_S}}{C^{MO}_{T_S}+S_{dyn}}$)

                                        \For{ n = 1 \textbf{to} $n_{s}$}
		                                  \State \textit{Update} the $St_{Task_{LA-n}}^{MO}$ $\And$ $d_{Task_{LA-n}}^{MO}$ 
		                    
		                                \EndFor
		                                
		                                \State /*Re-Mapping Checking*/
		                                \State  $Core_S$ $\gets$ $Core_{T_S}$, $Flag_{remap}$ $\gets$ 0
		                                \For{each core j in the cluster}

		                                    \If{$CF_{j}$ $<$ $\Gamma \times$ $CF_{Core_S}$ \textbf{and} free slack exists} 

		                                        \State $Core_S$ $\gets$ $C_j$, $Flag_{remap}$ $\gets$ 1;
		                    
		                                    \EndIf
		                    
		                                \EndFor
		                                \If{$Flag_{remap}$ == 1}    
		                                    Re-Map ($T_S$, $Core_S$);
		                    
		                                    \EndIf
		                            
		                             \EndIf
		                             
		                                \EndIf
		                  \For{\textbf{each} task \textit{i}}
		                    \If{$St_{i}^{MO}$ == $Time_{sys}$ or a task finishes its execution}
		                    
		                            
		                            \State DVFS(Ready and Running Tasks, Cores); //Update \par 
                					\hskip\algorithmicindent \hskip\algorithmicindent  cluster \textit{V-f} level (Algorithm 2)
		                    
		                    \EndIf
		                  \EndFor
		            
		        \EndProcedure

		    \end{algorithmic}
		}
	\end{threeparttable}
		
\end{algorithm}

If there is a dynamic slack during runtime, the algorithm selects the appropriate task to assign slack, which has more impact on instantaneous power consumption (lines~10-39). This dynamic slack is generated if a task finishes its execution before its defined WCET ($C_i^{LO}$ or $C_i^{HI}$ due to the system mode). In addition, since we use static scheduling of tasks for both modes and do not change the order of task execution in each core, there may be some idle time in a core that can be used. Therefore, if there is dynamic slack, we first compute the amount of available slack (lines~11-15). Hence, we have to consider the timing overheads of the scheduler and speed changing. Therefore we deduct these latencies from the slack to guarantee the deadline (line 17). Now, we select the appropriate task among $k$ tasks that can be released early due to the slack time after reclaimable slack (lines~18-23) based on the cost function (Eq.~\ref{eq:4}). Besides, Fig.~\ref{fig:Overal} details this process in the flow chart. After determining the appropriate task, according to the system mode situation, the frequency of the core to execute the appropriate task is obtained according to the amount of slack (line~25). Hence, the selected frequency must be rounded to the nearest \textit{V-f} level of the cluster that is greater. If there is at least one task between the generated slack and selected task, we change their start time. Therefore, their deadline would be changed (lines~26-28). Now, the re-mapping technique is applied if the core in which the task has been allocated, has a higher temperature than other cores (lines~29-37). As a result, it is possible to re-map the selected task to a core according to cost function (Eq.~\ref{eq:5}). As mentioned in Section~\ref{runtimephase}, we just re-map a task between the cores of each cluster. Further, the algorithm checks regularly that if a task is ready to start based on the static schedules, the \textit{V-f} level of the core that task has been mapped on it, is changed according to defined frequency scaling factor (lines~40-44). The detail is discussed in the following sub-section.

\subsection{Update \textit{V-f} Levels in Clustered Multi-Core Platform (DVFS governor)}
\label{governor}

After finishing a task execution, there might be a free slack or a task in the core queue that is ready to start its execution. Here, Algorithm~\ref{alg:DVFS} is executed to change the \textit{V-f} level if needed. As mentioned, all cores within a cluster operate at the same \textit{V-f} level in clustered multi-core platforms. 
Since the \textit{V-f} levels of both clusters are different, it is checked on which cluster the recently completed task was running (lines~2-4). Then, we check the assigned \textit{V-f} level of running or ready tasks on all cores of the cluster. Since all cores run with the same speed, we find the best frequency to set to the frequency cluster (lines~5-10). The reason for selecting the greatest minimum frequency is to ensure that all tasks finish their execution before their deadline. In the end, if the chosen frequency (\textit{SetFreq}) is different from cluster frequency, we change the speed of the cluster by assigning the new speed to one core of the cluster by using $\langle$cpufreq-set$\rangle$ program (lines 11-13). It should be mentioned that by changing the frequency of a cluster, its voltage will be changed automatically based on the table setting of the kernel. 

\begin{algorithm}[t]
    \caption{DVFS governor}
    \label{alg:DVFS}
    \begin{threeparttable}
		{ \small
				
			\begin{algorithmic}[1]
                \renewcommand{\algorithmicrequire}{\textbf{Input:}}
	        	\renewcommand{\algorithmicensure}{\textbf{Output:}}
					
                \Function {DVFS}{ Ready and Running Tasks, Cores}

                \If{\textit{$Core_{Task_i}$} $\leq$ 3} \quad $C_{ID}$ = 0; //Cluster with LITTLE \par 
                \hskip\algorithmicindent \hskip\algorithmicindent
                \hskip\algorithmicindent \hskip\algorithmicindent
                \hskip\algorithmicindent \hskip\algorithmicindent 
                \hskip\algorithmicindent \hskip\algorithmicindent 
                \hskip\algorithmicindent \hskip\algorithmicindent cores
                \Else \quad $C_{ID}$ = 4; //Cluster with big cores
                \EndIf
 
                \State SetFreq = 0;
                
                \For{\textit{c} = $C_{ID}$ \textbf{to} $C_{ID}$+3}
                
                   \If{\textit{SetFreq} $<$ $Freq_{Run/Ready Task on Core_{C_{ID}}}^{MO}$}
		                    
	                   \State SetFreq = $Freq_{Run/Ready Task on Core_{C_{ID}}}^{MO}$;
		                    
		            \EndIf
                
                \EndFor
                
                 \If{\textit{SetFreq} $!=$ $Freq_{Cluster}$}
		                    
	                \State \texttt{cpufreq-set}  -\texttt{c}  $Core_{Task_i}$  -\texttt{f}  \textit{SetFreq}
		                    
		         \EndIf
		         
               \EndFunction 
      
		    \end{algorithmic}
		}
	\end{threeparttable}
\end{algorithm}


\begin{table*}[t]
    \centering
    \caption{Run-time scheduler cache misses report}
    \vspace{-5pt}
    \label{reportcashehm}
    \begin{center}
        \begin{tabular}{|c|c|c|c|c|c|c|}
        \hline
        \textbf{ } & \multicolumn{3}{|c|}{\textbf{Look-Ahead Unit} } &  \multicolumn{3}{|c|}{\textbf{Re-mapping Unit} }\\
        \hline
        \textbf{} & \textbf{Cortex A7} &  \textbf{Cortex A15} &  \textbf{Intel Core i5} &  \textbf{Cortex A7} &  \textbf{Cortex A15} &  \textbf{Intel Core i5}\\
        
        \hline
        \textbf{L1 Data Read Miss}& 3.318\% & 3.318\% & 2.946\% & 0.163\% & 0.163\% & 0.303\% \\
        
        \hline
        \textbf{L1 Data Write Miss}& 3.584\% & 3.584\% & 2.168\% & 0.352\% & 0.352\% & 0.764\%  \\
        \hline
        \textbf{LL Data Read Miss} & 0.076\% & 0.081\% & 0.0\% & 0.028\% & 0.034\% & 0.0\% \\
        \hline
        \textbf{LL Data Write Miss} & 0.063\% & 0.0\% & 0.0\% & 0.036\% & 0.0\% & 0.0\% \\
        \hline
        \end{tabular}
    \end{center}
    \vspace{-6pt}
\end{table*}


\section{Run-Time Scheduler Algorithm Optimization--Analysis and Implementation}
\label{OVERHEAD}

In the presented approach of our previous paper~\cite{Ranjbar2019}, the timing overheads of run-time scheduling and changing the frequency have not been considered. However, these overheads can have a profound impact on power-aware run-time scheduling of tasks and must be considered in the respective analysis. Neglecting them may lead to missing deadlines for MC tasks, which may cause catastrophic consequences. Two sources of generating overheads that deal with the online scheduler are the Look-Ahead Unit to select the appropriate tasks and the Re-mapping Unit to find the appropriate core. In the following, we use the online scheduler phrase for both units to make it easy to follow.
The other source of causing overhead is the DVFS governor Unit for changing frequency during runtime. Now, we first analyze the scheduler function from the timing overhead aspect. Then, we focus on optimizing the code and reducing the overheads.

In order to evaluate the scheduler and analyze the overhead on a real platform, we first convert Matlab code to C code. Then, we detect the main parts of the code, which have more latency and attempt to optimize it.
To analyze the main functions, we first get a strict upper bound of the latency in different parts of the online scheduler on a real platform. We use the Kcashegrind tool~\cite{kcache} to measure the worst-case time. Kcashegrind is a visualization tool that uses a technique called profiling, which gives you the time distribution among the scheduler code at runtime. Now, we focus on the functions code and its timing analysis and endeavor to reduce the estimation cycles and the delay caused by cache misses in shared cache levels. 
Both Look-Ahead and Re-mapping unites in the online scheduler are called frequently during runtime, and the apparent improvement and optimization should be performed. The run-time phase of Fig.~\ref{fig:Overal} shows the flow chart of these two units in detail. As shown in this figure, some functions play a critical role in timing overhead of power-aware run-time scheduler, which is indicated by white color. As discussed in the previous section, the Look-Ahead unit chooses $k$ tasks after generated dynamic slack and find a task that has the most effect on peak power and maximum temperature. Checking the $k$ tasks is done in a for-loop, in which each task is investigated that can release early to use the dynamic slack. Therefore, all predecessors of it must be checked whether they can finish their execution soon or not. Investigating the execution status of all predecessors need more cycles to be done and then causes latency and more cache misses. Therefore, introducing an entity that shows the estimated finish time of a task would be useful, and instead of checking the status of all predecessors, just that entity can be checked.
Besides, due to the having different \textit{V-f} levels, we must ensure that the dynamic slack is large enough to include the timing overhead of changing the \textit{V-f} level. This check prevents the over-calling of the Re-mapping Unit. In addition, there are two functions of calculating the costs, in which there are some math calculations with high timing overhead. Hence, optimizing these calculations by pre-defining them to avoid dynamic memory allocation during computation would help reduce timing overhead.

From the perspective of cache hit/miss, one of the ideas is to optimize the code to reduce the estimation cycles in the online scheduler by focusing on calculations and memory access latency. There are some tips to optimize C/C++ code to run it faster; reducing functions calls and the number of function parameters, how to define variables and objects, how to use operators, using prefix instead of postfix in objects, avoiding unnecessary data initialization and so many other techniques that we must use for code optimization.
Apart from using these techniques, due to data access latency, we have effective timing overhead in the online scheduler. We optimize code by changing the representation of the data structure manipulated by the algorithms. We have defined two types of task classes: 1) defining a task class that uses vectors in the class for each task entity, 2) defining a task class with vectors of task class in the number of tasks. Each has its advantages and disadvantages under certain circumstances. However, due to the checking of limited tasks ($k$) in the run-time scheduler, the use of the second task class has less timing overhead, and cache misses.
As a result, Table~\ref{reportcashehm} shows the percent of cache (L1 and LL (last level)) read and write misses for Look-Ahead and Re-mapping Units after optimization on three different platforms, Cortex A7, Cortex A15, and Intel Core i5. As mentioned in previous sections, most of the embedded systems use ARM processors, not Intel. Therefore, we target the ARM processors, such as the ODROID board. However, this table shows that we have less than 3.584$\%$ and 0.081\% cache L1 and LL data misses in ARM processors, respectively, which are admissible compared to all cache misses and also in comparison with Intel processor that has less cache misses.

\section{Evaluation}
\label{lemma:Results}

\subsection{Experimental Setup}

\subsubsection{Hardware Platform}

To evaluate our system, we conducted experiments on the ODROID~XU3/XU4 board powered by ARM, which has big.LITTLE architecture, four big (Cortex A15), and four LITTLE (Cortex A7) cores. As explained in Section~\ref{SystemModel}, the ODROID~XU3 board supports DVFS and can operate at 13 different \textit{V-f} levels between $[0.9V,200MHz]$ and $[1.3V,1.4GHz]$ on LITTLE cores, while the last four frequency levels have the same voltage levels and 19 different \textit{V-f} levels between $[0.9V,200MHz]$ and $[1.3625V,2GHz]$ on big cores. Therefore, the effect of changing \textit{V-f} levels is done by scaling the frequency within the range of available levels. 

\vspace{3pt}

\subsubsection{Task Set Generation}

In the experiments, we use random applications (task graphs) generated by the tool in~\cite{Medina2018}. An example of a real-life application is already given in the motivational example. In these applications, there are four basic parameters, \textit{c} (number of cores), \textit{U} (system utilization), \textit{d} (outgoing edge percentage) and \textit{n} (number of tasks), which are presented in Table~\ref{experimenrs}. \textit{d} represents the probability of having outward edges from one task to the others. In addition, $U/c$ is the normalized system utilization that refers to both LC and HC tasks with their predefined $C^{HI}$. As the results are presented in both simulation and real platform (with eight cores), we show the results with 16 cores in simulation in addition to 2, 4 and 8 cores.
We provide different configurations by changing the value of these parameters for different scenarios used in the experiments. 

\vspace{3pt}

\subsubsection{Tasks' Power Consumption}

In order to have a realistic possible range of power values, we run several embedded benchmarks from MiBench suite~\cite{Mibench}, e.g., automotive, network and Telecomm benchmarks on two configurations, ARM Cortex A7 and A15 on the ODROID~XU3 with maximum frequency and read data from power sensors on the board. Hence, since the DVFS is applied to the whole processor, the power consumption at other lower frequencies can be obtained using Eq.~\ref{eq:rhomax} in Section~\ref{Powermodel} by considering frequency scaling~\cite{Salehi2015}. 
In addition, we examined different scenarios of activating one core to all cores by running different benchmarks. We run each benchmark 1000 times and report the maximum value of power consumption. We select the maximum power of tasks in the range of these minimum and maximum values in our experiments, which is [484,~940]\textit{m}\textit{W} in Cortex A7 and [3.891,~7.622]\textit{W} in Cortex A15. The power that the tasks may consume is generated randomly following the normal distribution within this range. Besides, we consider the power consumption of the system as the sum of the power consumption of all cores~\cite{Munawar2014}.

\vspace{3pt}

\subsubsection{Thermal Analysis}

As presented in the proposed method section, we assume that our approach does not have to probe the core temperature to make a decision. Therefore, during the scheduling of the tasks, the power values of cores depending on the running tasks are recorded. In addition, for validating on the real platform, since there are just temperature sensors for big cores on the ODROID~XU3, the HOTSPOT tool~\cite{Skadran2006} is used to obtain the core temperature throughout the execution for the specific floorplan and configuration platform which we use. For the configuration file, we use the parameters reported in~\cite{ARM2018}, which is for ARM big.LITTLE processors. The ARM core (A7) has an area of 0.45 $mm^2$ in our experiments reported by the ARM company.

\begin{table}[t]
    \centering
    \caption{Experiment Configurations}
    \vspace{-4pt}
    \label{experimenrs}
    \begin{center}
        \begin{tabular}{|c|c|c|c|c|}
        \hline
        \textbf{Param.} & \textbf{ \thead{Varying \\c}} &  \textbf{\thead{Varying \\U/c}} & \textbf{\thead{Varying \\n}} & \textbf{\thead{Varying \\d}} \\
        \hline
        $c$ (\#core) & \centering 2, 4, 8, 16 & \centering 8 & \centering 8 & 8 \\
        \hline
        $U/c$ (utilization) & [0.5, 0.75] & [0, 1]& [0.5, 0.75] & [0.5, 0.75]  \\
        \hline
        \thead{$d$ \\(edge percentage)}& 10\% &  10\% & 10\% & \thead{1\%, 10\%, \\ 20\%} \\
        \hline
        \thead{$n$\\(\#task)} & \centering 50 & \centering 50 & \thead{30, 40, \\50, 80} & 50 \\
        \hline
        \end{tabular}
    \end{center}
    \vspace{-4pt}
\end{table}

\vspace{3pt}
 
\subsubsection{Comparison}

In this paper, we analyze our modified proposed method and compare against~\cite{Medina2018,DZhu2003}, 
and the previous work,~\cite{Ranjbar2019}. The work~\cite{Medina2018} proposes an offline scheduling algorithm for an MC system where most of the LC tasks are not dropped in the HI mode to improve the QoS of the system. However, they ignore the peak power and temperature aspect of the system. Additionally, researchers in~\cite{DZhu2003} suggest an online energy minimization algorithm for hard real-time systems where they use the dynamic slack just for the immediately available task to decrease the \textit{V-f} level. We compare with the method of this paper by considering the latency to have a fair comparison.

\subsection{Experimental Results, Observation and Discussion}


\subsubsection{\textbf{Analyzing the relevance between a core temperature and energy consumption}}

\begin{figure}[t]
\centering
\includegraphics[width=0.93\columnwidth]{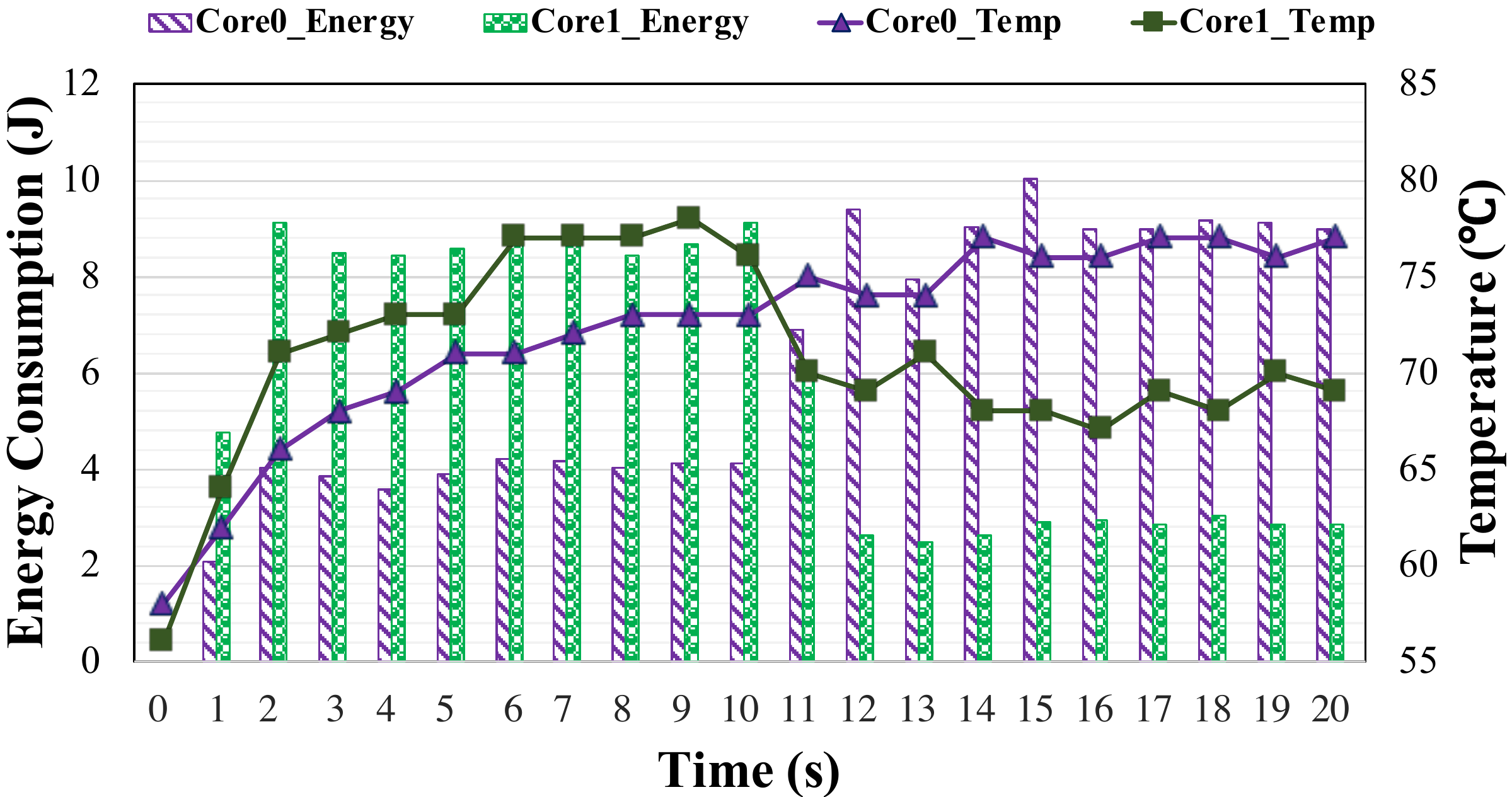}
\caption{The relevance between a core temperature and accumulated energy consumption.}
\label{fig:EnergyTemp}
\vspace{-5pt}
\end{figure}

At first, we represent the relevance between core temperature and its accumulated energy consumption. In Section~\ref{CoreRemapT}, our algorithm was based on the assumption that a core tends to have a lower temperature when its accumulated energy is less than the other cores. Fig.~\ref{fig:EnergyTemp} studies the validity of the assumption. Since we do not have a power sensor for each big core on the ODROID~XU3, we run the same task on all the big cores to have the same power consumption. This task is executed several times periodically in cores with different execution times. Therefore, we have different energy consumption in each period of cores. After finishing the execution of the task on each core, the core goes to sleep until the end of the task's period. 
Fig.~\ref{fig:EnergyTemp} shows energy consumption and the temperature of two big cores during the time for a window of energy monitoring equal to two seconds. In this figure, first, the task runs with larger execution time on Core1 in comparison to Core0. Thus, the temperature of Core1 rises more rapidly than Core0. After 10$s$, the accumulated energy of Core1 is reduced, and Core0 is increased. As shown, the Core0 that has more energy consumption tends to have a higher temperature.

\begin{figure*}[t]
	\centering
	\begin{minipage}[t]{1\linewidth}
		\centering
        \subfloat[Normalized Peak Power]{\includegraphics[width=0.32\linewidth]{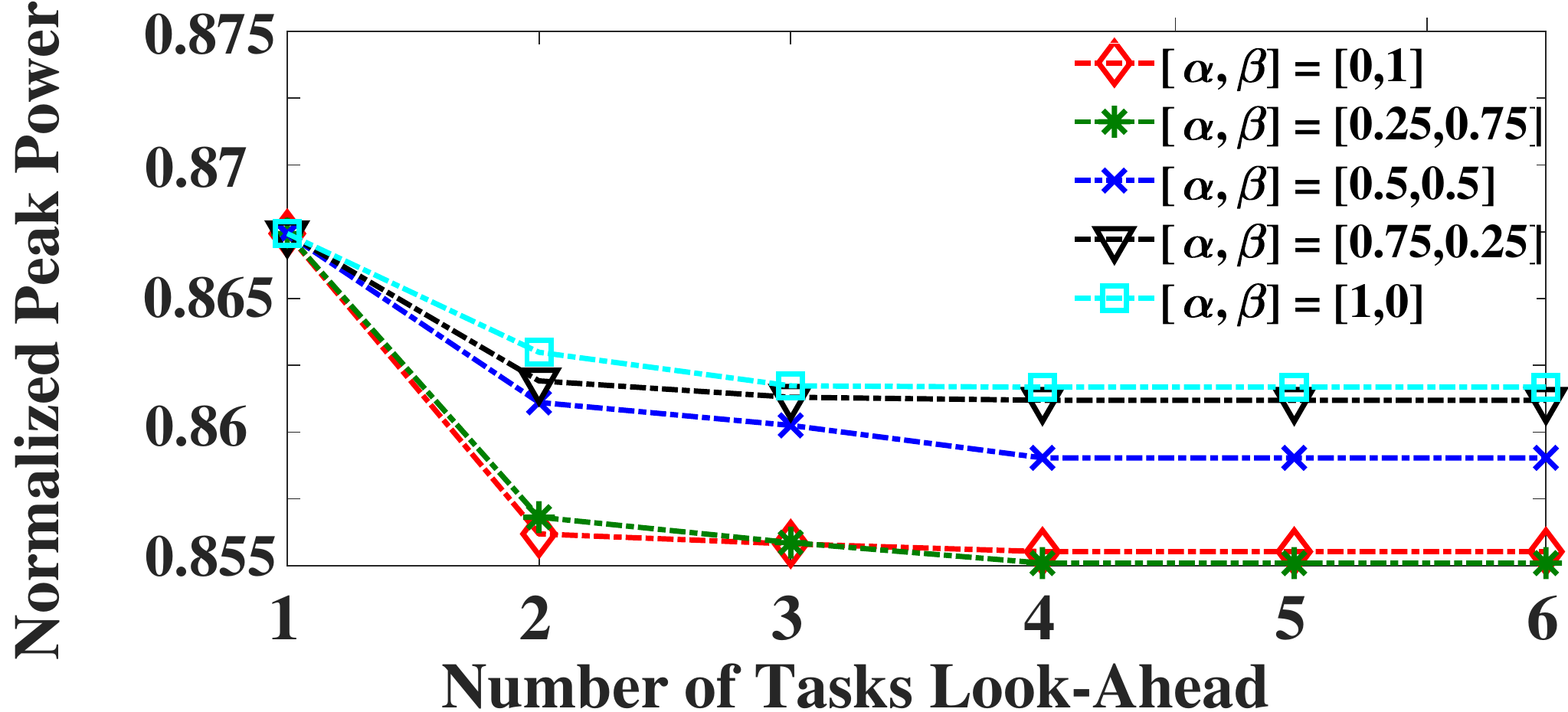}\label{fig:fig4a}}
        \hfil
        \subfloat[Normalized Energy]{\includegraphics[width=0.32\linewidth]{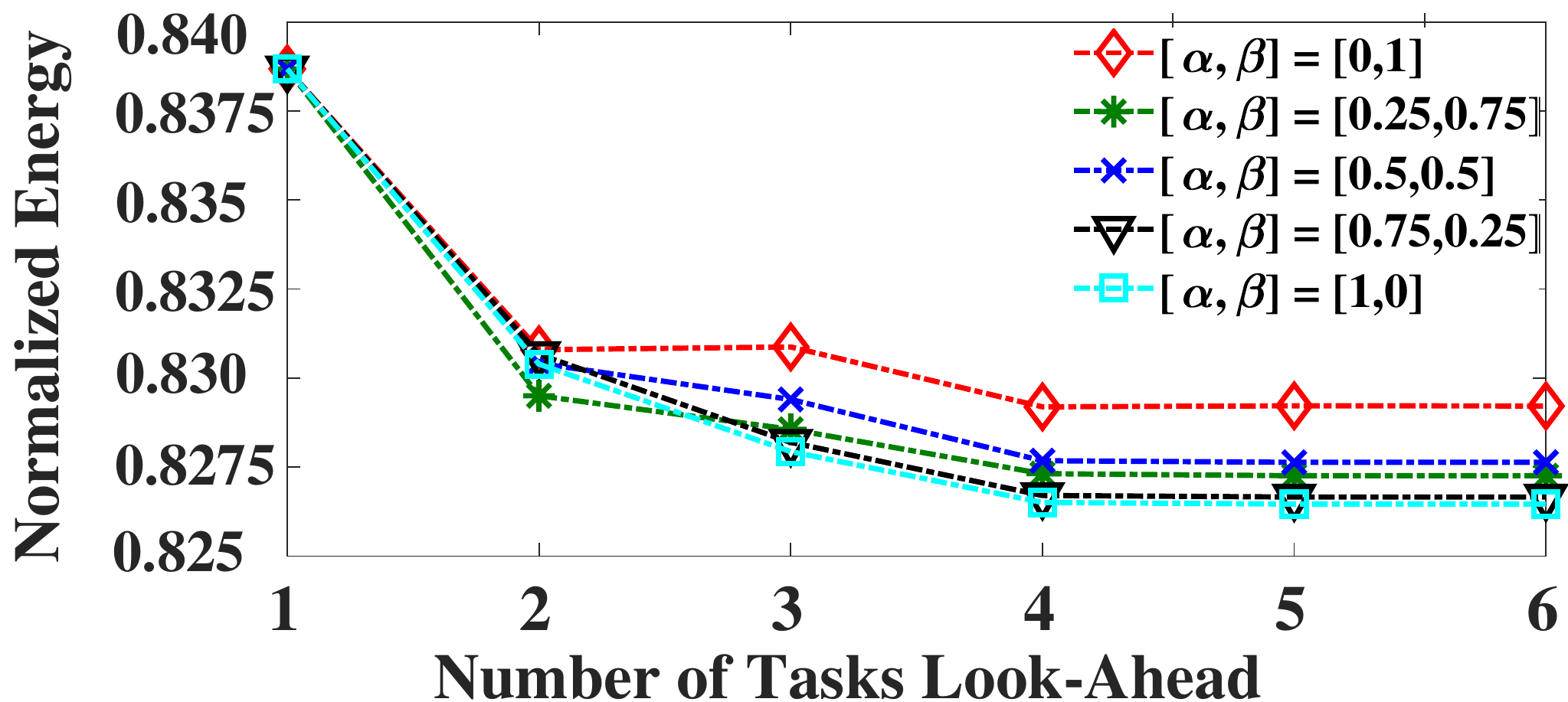}\label{fig:fig4b}}
        \hfil
        \subfloat[Normalized Peak Temperature]{\includegraphics[width=0.32\linewidth]{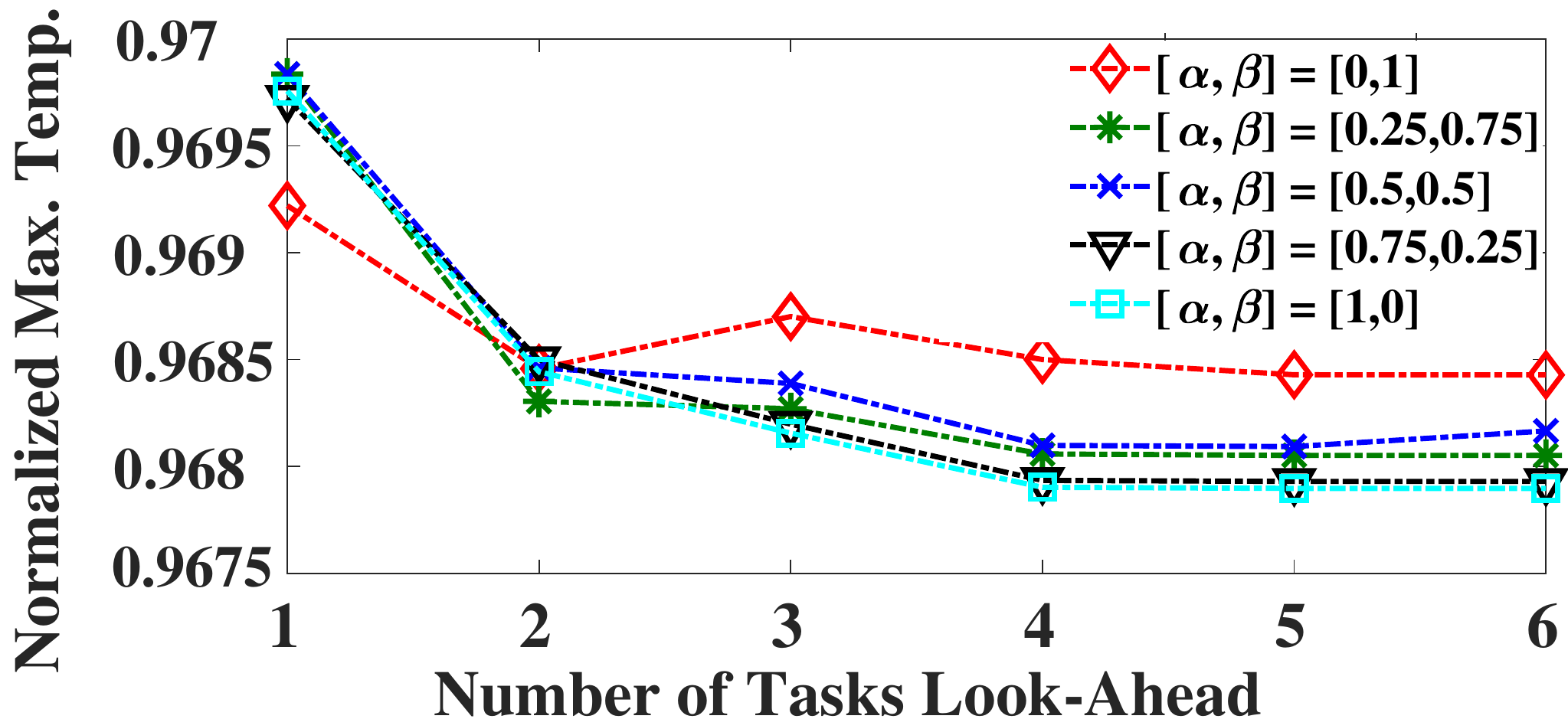}\label{fig:fig4c}}
		\caption{Impact of varying $\alpha$ and $\beta$ on peak power, energy and max. temperature.}
		\label{fig:alphabeta}
	\end{minipage}
\end{figure*}

\vspace{5pt}

\subsubsection{\textbf{The effect of varying $\langle \alpha, \beta \rangle$}}

Now, we evaluate the results for different values of $\alpha$ and $\beta$ in Eq. \ref{eq:4}. The experiments are carried out for a system with $c = 8$, $U/c \in [0.5, 0.75]$, $d = 1\%$ and $n = 30$. The average results (Fig.~\ref{fig:alphabeta}) are obtained for a set of 100 task graphs with different $\langle \alpha, \beta \rangle$ = $\langle 0, 1 \rangle$, $\langle 0.25, 0.75 \rangle$, $\langle 0.5, 0.5 \rangle$, $\langle 0.75, 0.25 \rangle$ and $\langle 1, 0 \rangle$. The results are normalized to~\cite{Medina2018}. In this section, to show the effect of varying these two parameters, tasks are executed with their actual execution time, and task re-mapping is not exploited. It can be seen that, in every case, utilizing our approach would lead to a system with lower peak power, energy as well as peak temperature. Besides, the expected effect of varying $\langle \alpha, \beta \rangle$ is confirmed in the experiments. For example, the average normalized peak power is progressively reduced when $\beta$ increases from 0 to 1, as presented in Fig.~\ref{fig:fig4a}. Similarly, in Fig.~\ref{fig:fig4b}, the higher the $\alpha$, the lower the energy consumption and peak temperature. Finally, as the algorithm looks further ahead in the future to find the best tasks to assign the dynamic slack, the results are generally getting better, up to $1.25\%$, $1.25\%$, and $0.22\%$ more reduction in peak power, energy and peak temperature. It is worth noting that, in this experiment, we intentionally disable the task re-mapping technique to ensure that the effect of $\langle \alpha, \beta \rangle$ is not skewed by another optimization.

For the other experiments in the paper, we consider $\langle \alpha, \beta \rangle = \langle 0.5,0.5 \rangle$ that balances both peak power and temperature average reduction in comparison with other values of $\langle \alpha, \beta \rangle$.

\vspace{5pt}

\subsubsection{\textbf{The optimum number of tasks to look ahead and the effect of task re-mapping}}

\begin{figure}[t]
\centering
\subfloat[Peak Power]{\includegraphics[width=0.32\columnwidth]{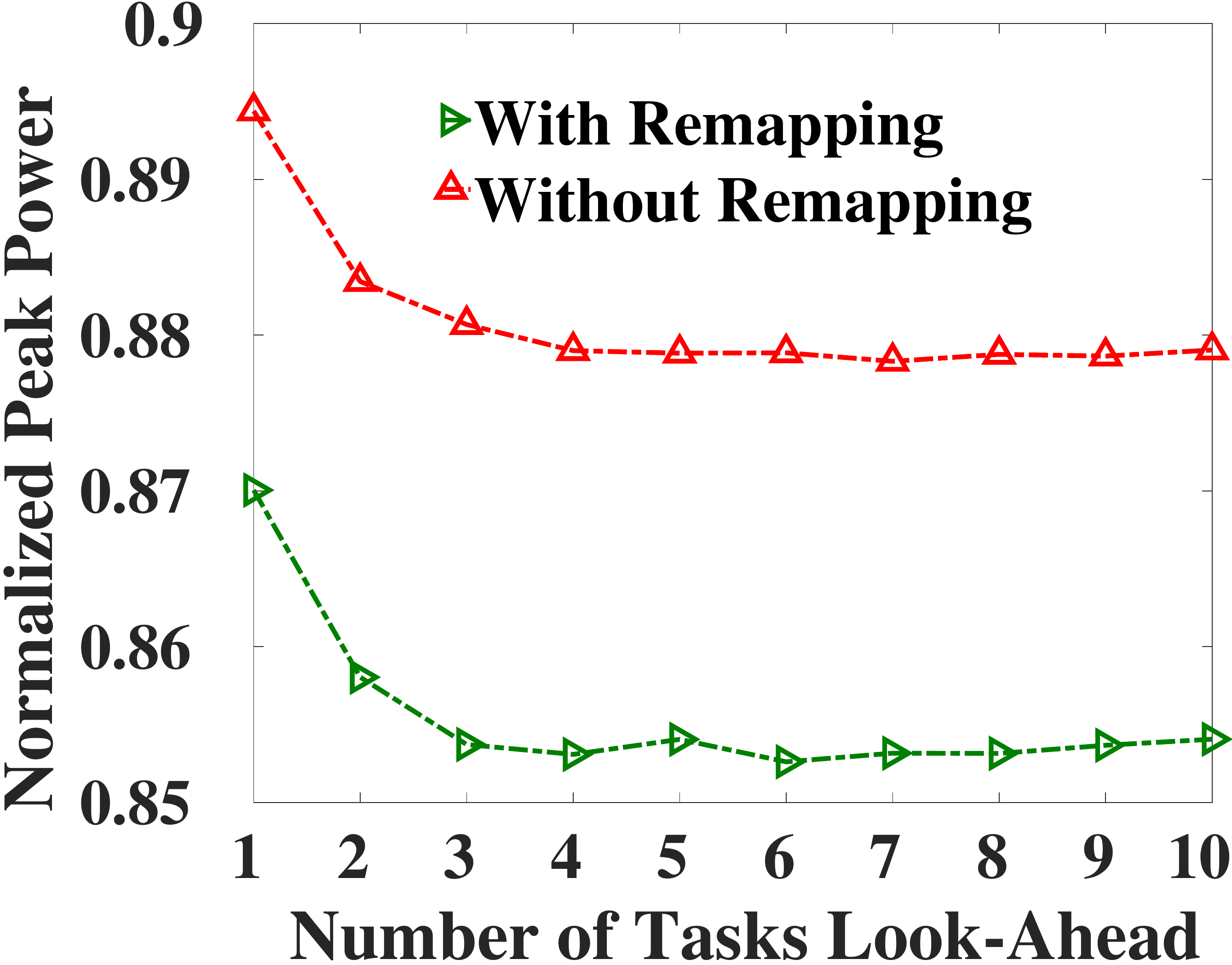}\label{fig:Allpower}}
\hfil
\subfloat[Energy]{\includegraphics[width=0.32\columnwidth]{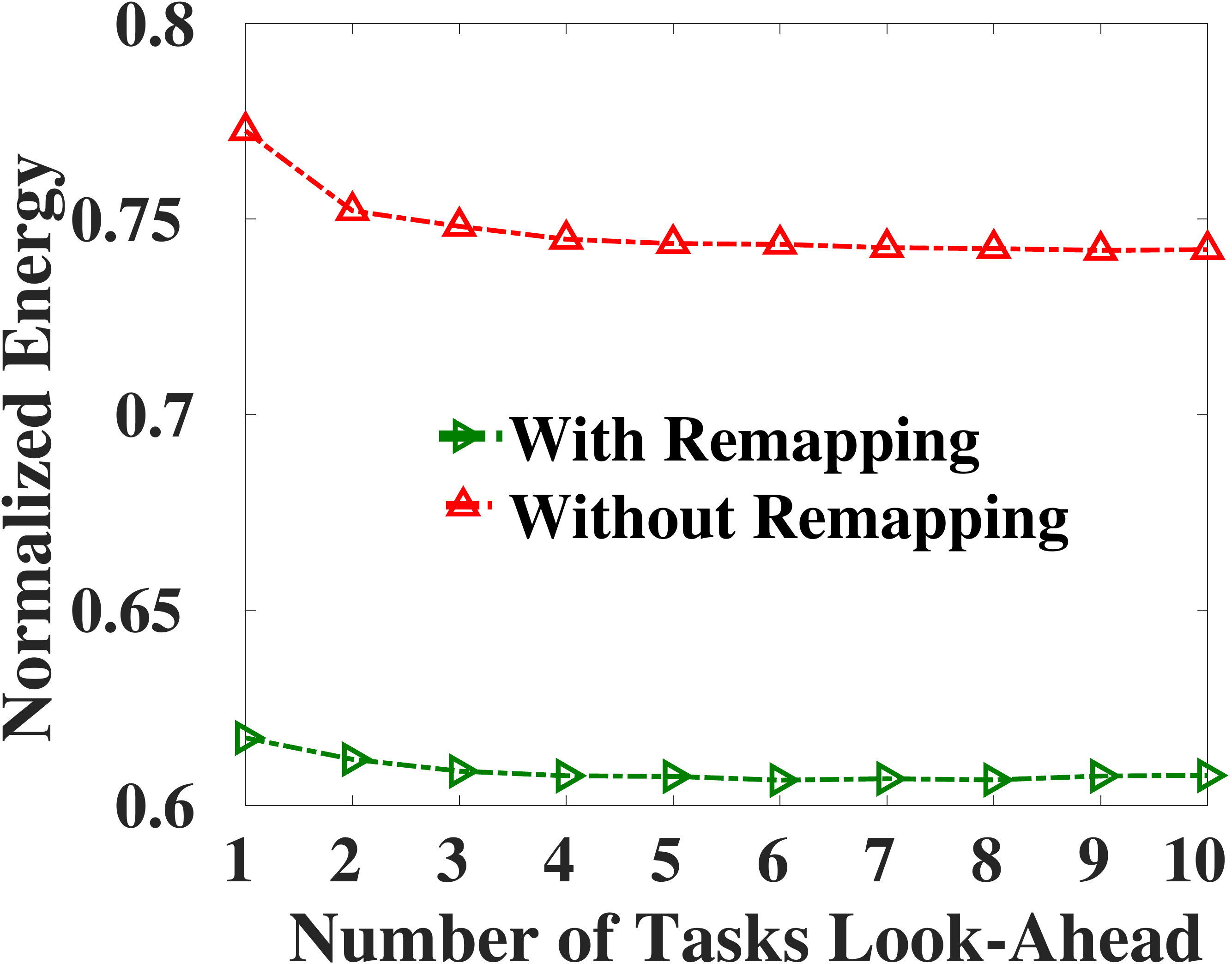}\label{fig:allenergy}}
\hfil
\subfloat[Max. Temperature]{\includegraphics[width=0.32\columnwidth]{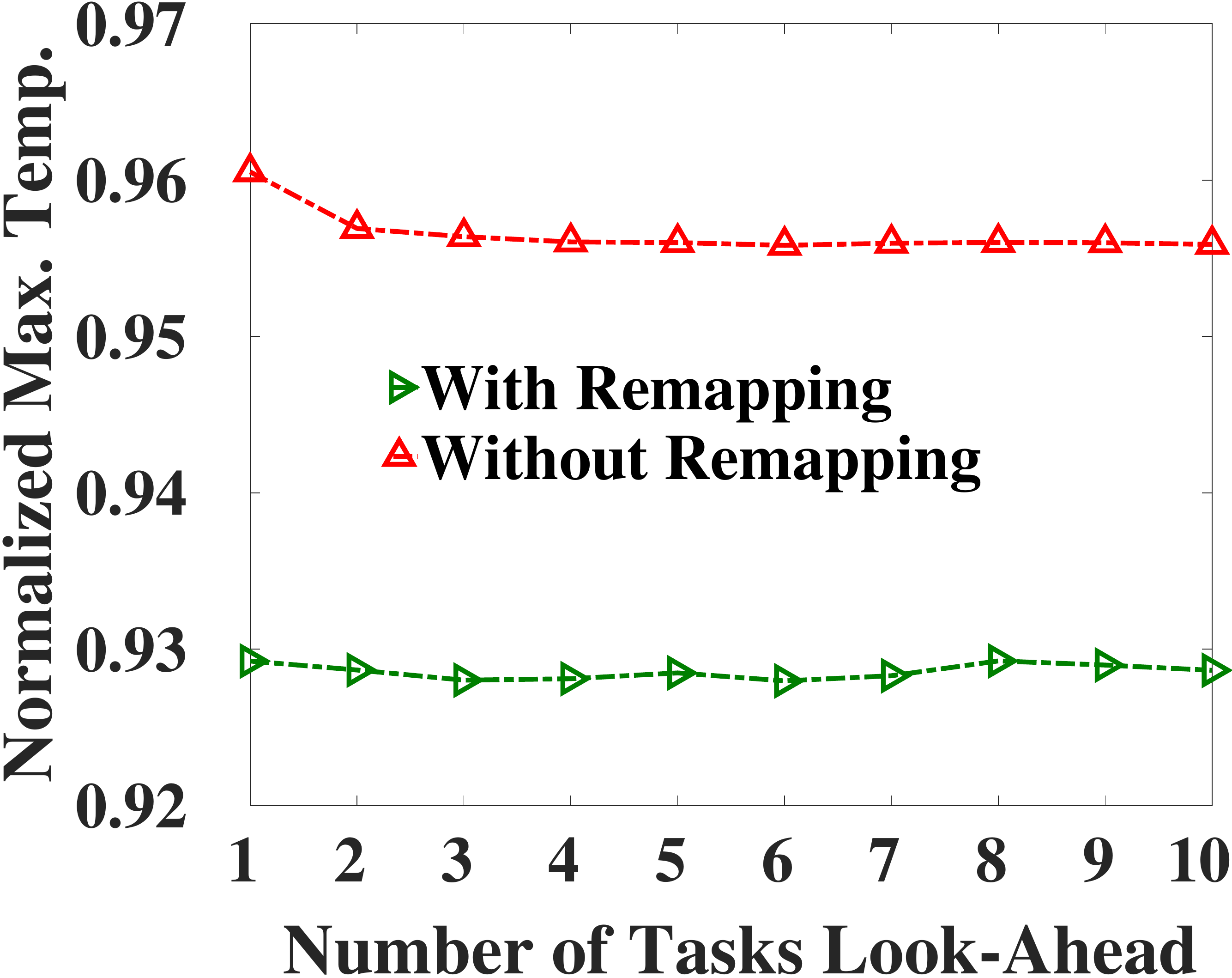}\label{fig:alltemp}}

\caption{Normalized improvement in peak power, energy and max. temperature for all scenarios.}
\label{fig:Allimp}
\vspace{-6pt}
\end{figure}

In this subsection, we analyze the optimum number of tasks to look ahead ($k$) by evaluating the respective average quality of results without considering the overheads. The number of look-ahead tasks is varied from 1 to 10. The results presented in Fig.~\ref{fig:Allimp} are obtained from some scenarios of changing parameters in Table~\ref{experimenrs} with running on a homogeneous multi-core system. 
As a result, looking 4 tasks ahead provides a significant reduction in peak power and also in maximum temperature and energy consumption with and without task re-mapping. 
Hence, looking four tasks ahead is the average result of varying all parameters. The detail of finding the optimum $k$ by varying the properties of tasks is discussed in Appendix B. 
Besides, when task re-mapping is used, the temperature, on average, is reduced by $2.7\%$ compared to the case where task-remapping is disabled. In general, by looking ahead 4 tasks and enabling task re-mapping, the proposed method reduces the peak power, energy consumption, and maximum temperature on average by 14.6\%, 39\%, and 7.1\%, respectively compared to~\cite{Medina2018} and 4.2\%, 16\%, and 3.1\%, respectively compared to~\cite{DZhu2003}.

\vspace{5pt}

\subsubsection{\textbf{The analysis of scheduler timings overhead on different real platforms}}

\begin{figure}[t]
\centering
\includegraphics[width=1\columnwidth]{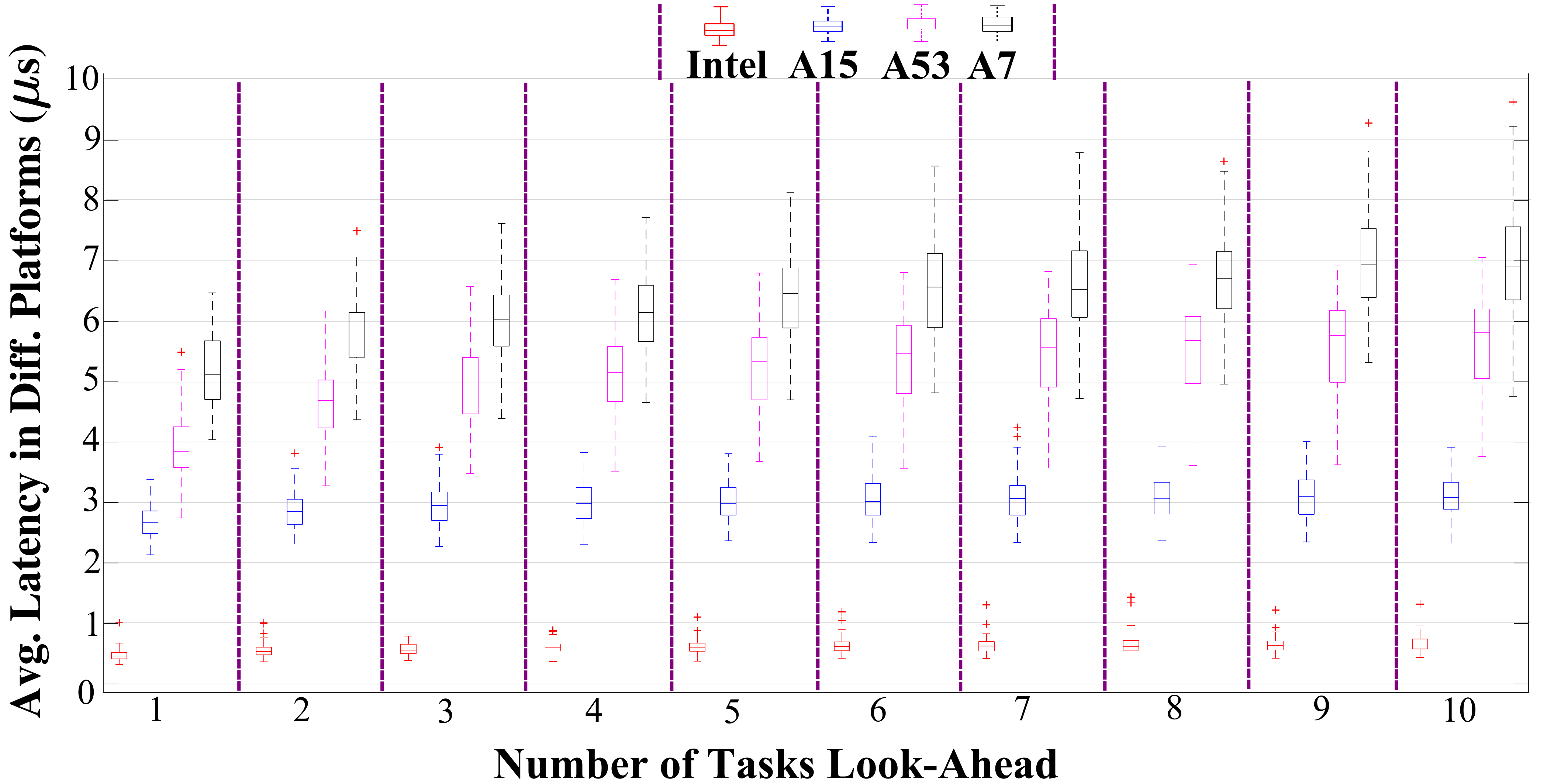}
\caption{Analyzing the timing overhead of run-time scheduler on four platforms.}
\label{fig:platforms}
\vspace{-5pt}
\end{figure}

To investigate the timing overhead of the proposed run-time scheduler, we analyze it on three real platforms, Intel Core i5, ARM big core (A15), and ARM LITTLE core (A7) on ODROID~XU3/4 and ARM core (A53) in Xilinx Zynq UltraScale+ MPSoC board. Fig.~\ref{fig:platforms} shows the overheads in each platform for different numbers of tasks looking ahead. Each boxplot shows the average latency for normal values of parameters in Table.~\ref{experimenrs} (d= 10\%, c= 8, U/c= [0.5, 0.75], n= 80) with 100 task graphs. 
The following observation can be seen from the figure. First of all, the run-time timing overhead in the Intel platform is extremely small as compared to ARM processors. In the second observation, as the number of tasks look-ahead is increased, the latency is increased in all ARM processors. However, this latency increase is more evident in the A7 processor, while it is almost constant after looking seven tasks ahead in the A53 processor and four tasks ahead in the A15 processor. 
Furthermore, since big (A15) cores have high performance as compared to LITTLE (A7) cores, this timing overhead would be less. However, since a platform has lower performance, the range of latency between the minimum and maximum value is more significant. This fact is due to its lower performance and access to the memory and cache miss/hit. 

To have a real implementation of our proposed method, we obtain the observed worst-case timing overhead of run-time scheduler, in which the appropriate task is selected for slack assignment and also a proper core for the re-mapping. 
Since many embedded systems use ARM processors, we evaluate our method on the ARM processor. 
We analyze the overhead on a LITTLE core of the ODROID XU4 platform. We examine both the Look-Ahead Unit and Re-mapping Unit in the run-time scheduler, separately and obtain the maximum observed timing overhead.
To determine this overhead, we ran several applications (200 task graphs) with their various inputs on ARM LITTLE Core (A7). Based on these overheads, we set the observed worst-case of the Look-Ahead Unit to the maximum value. In addition, the scheduler in Re-mapping Unit checks other cores, whether it is possible to re-map a task for thermal management. In the worst-case, all cores are checked. Therefore, to have a close to accurate observed worst-case timing overhead of the Re-mapping Unit, we obtain the worst timing overhead for each core and multiply it by the number of cores. Based on our observation and this measurement, the maximum timing overhead for Look-Ahead Unit is 56.417$\mu$s, and for Re-mapping Unit per core is 64.54$\mu$s. In order to ensure that the timing guarantees provided by the static schedule are not violated, we deduct these overheads from slack before assigning it to an appropriate task.

\vspace{5pt}

\subsubsection{\textbf{The latency of changing frequency in real platform}}
\label{FreqChange}

The main unit, DVFS governor, adjusts the frequency, which has a significant timing overhead.
The ODROID~XU3/4 board has a frequency range of $\langle0.2,1.4\rangle$ GHz for LITTLE cores and $\langle0.2,2\rangle$ GHz for big cores, with the step of 0.1 GHz. We change the frequency by using $\langle$cpufreq-set$\rangle$ program in two scenarios of scaling-down and scaling-up. Hence, the voltage is adjusted automatically according to the selected frequency. The maximum latency for all scenarios is at most 12.025\textit{ms}. Besides we observe in our experiments that the latency of scaling-down transition is 342$\mu$s less than the scaling-up transition, on average. 
Regardless of the frequency scaling-down or up, we consider the latency of changing \textit{V-f} level to be equal to 12.025\textit{ms}. Due to this timing overhead, we deduct this latency from available dynamic slack before assigning it to a task to guarantee the correct execution of tasks before their deadlines. Since the re-mapping latency is 3.75$m$s~\cite{Basireddy2019} in the worst-case scenario and remapping is done in parallel with changing the frequency, it has no impact on the overall deadline.

\vspace{5pt}

\subsubsection{\textbf{The effect of latency on system schedulability}}

\begin{figure}[t]
\centering
\includegraphics[width=1\columnwidth]{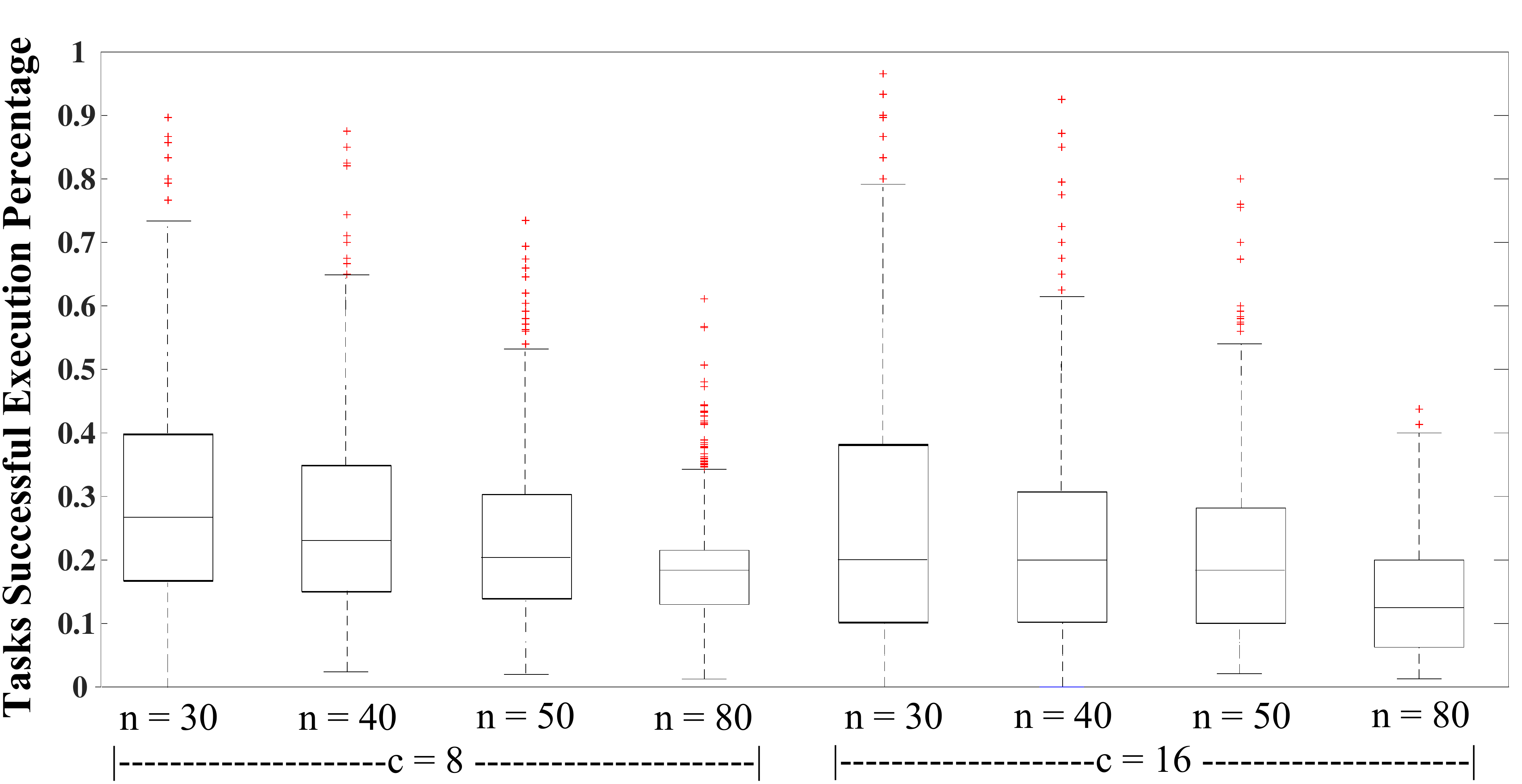}
\caption{Percent of successful executed tasks before their deadline in task graph based on the proposed method in~\cite{Ranjbar2019} without considering timing overheads.}
\label{fig:taskschdule}
\end{figure}

As discussed, considering the latencies of the run-time scheduler and changing frequency are critical in analyzing the system, which has not been considered in our previous work~\cite{Ranjbar2019}. If these timing overheads are not studied, it may cause deadline miss of tasks and then catastrophic consequence. Our proposed method's effectiveness depends on the available slacks at runtime and the possibility of assigning them to the tasks. Therefore, if the latencies are not properly accounted for, some tasks may not be executed successfully before their deadline. 
Fig.~\ref{fig:taskschdule} shows the percentage of successfully executed task sets before their deadlines during run-time phase in different scenarios in the method of~\cite{Ranjbar2019}. The results are obtained for the normal scenario of some parameters (d=~10\%, U/c=~[0.5,0.75], c = 8, 16 and n = 30, 40, 50 and 80) and 1000 task graphs for each scenario. 
In this figure, we observe that as the number of tasks in the system with the same number of cores is increased, fewer task sets can be scheduled and meet their deadline. When there are more tasks in the system with the same \textit{U/c}, the dynamic slacks that are incurred when the tasks finish earlier than their WCETs are smaller. The reason is that as the expected execution times of the tasks are decreased, the absolute differences between their actual execution time and WCETs are inherently small. Therefore, the possibility of missing a deadline is increased, and fewer tasks would be executed successfully before their deadline. In addition, if the number of cores in the system is increased, more task sets miss their deadlines. Since the re-mapping technique is used to manage temperature, all cores are checked in the worst-case. Therefore, by increasing the number of cores, the timing overhead of selecting a proper core for task re-mapping is increased. Since this latency has not been considered in~\cite{Ranjbar2019} while a dynamic slack is assigned, using the re-mapping technique at runtime, may cause more deadline violation by increasing the number cores. In general, as shown in this figure, a high percentage of task sets miss their deadline, which is not acceptable in MC systems. Therefore, it is critical to consider timing overheads of run-time scheduler and changing frequency.

\begin{figure}[t]
	\centering
	\begin{minipage}[t]{1\linewidth}
		\centering
        \subfloat[Varying Number of Cores]{\includegraphics[width=0.49\columnwidth]{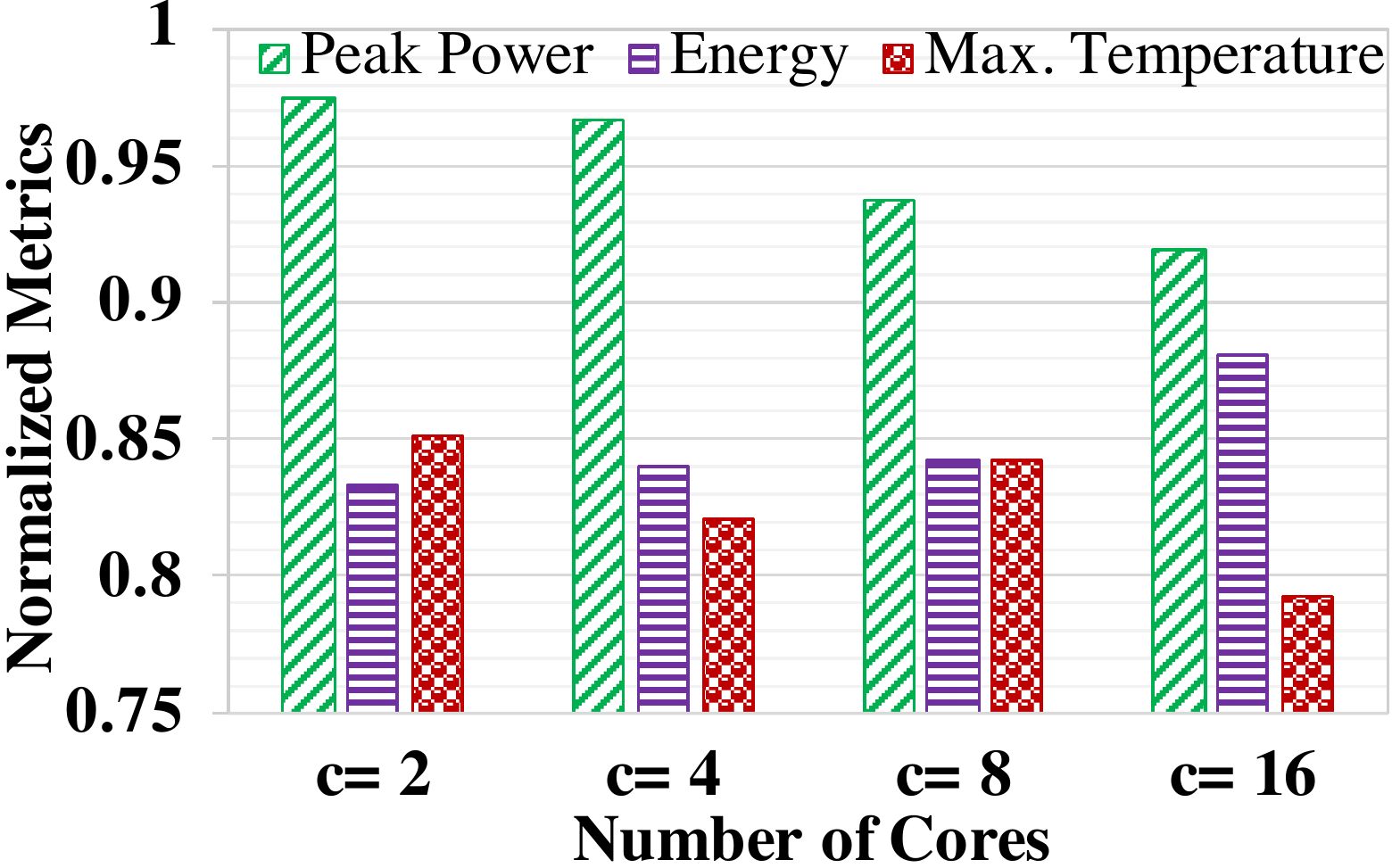}\label{fig:AllCores}}
        \hfil
        \subfloat[Varying Utilization Bound]{\includegraphics[width=0.47\columnwidth]{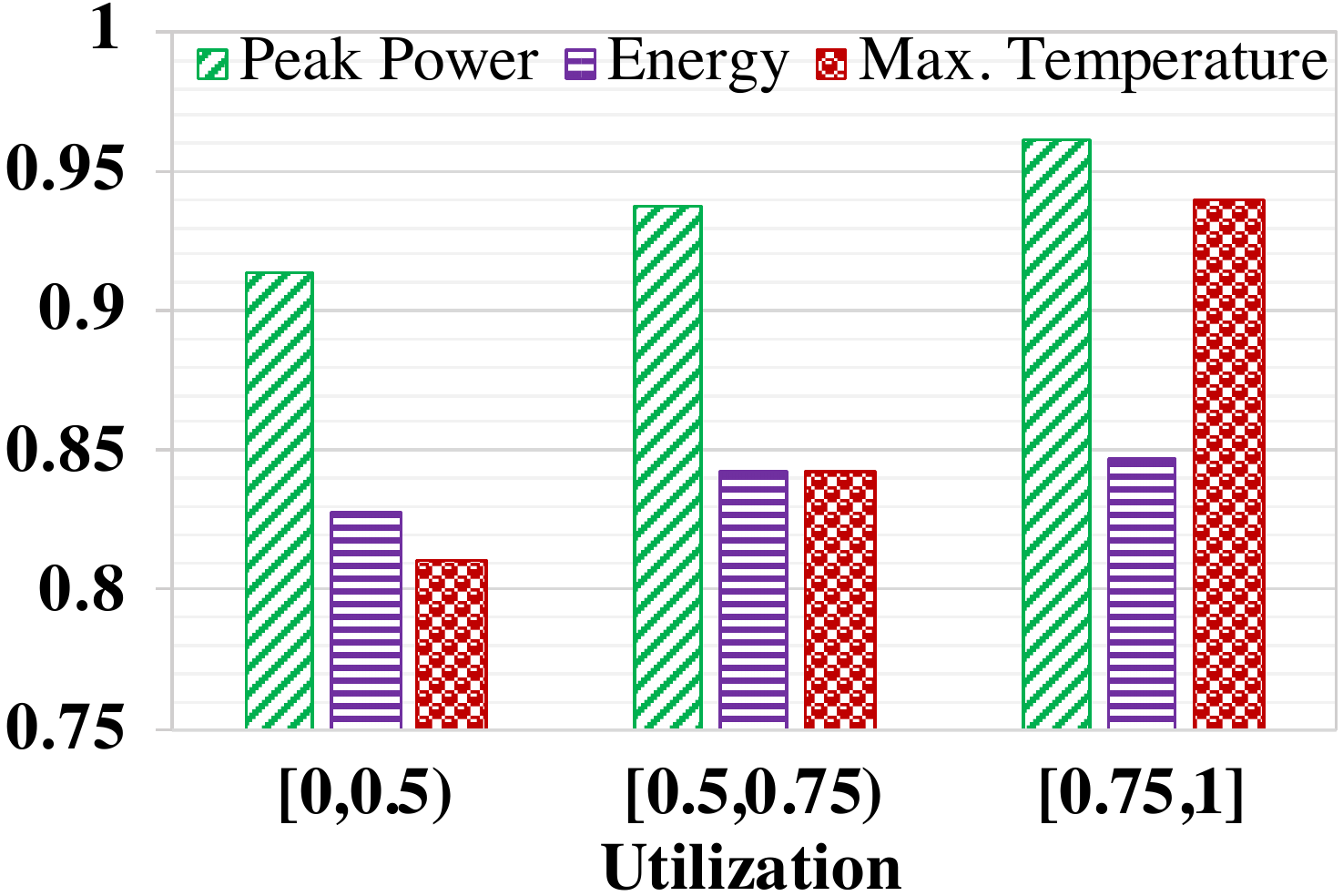}\label{fig:AllUti}}
        \hfil
        \subfloat[Varying Number of Tasks]{\includegraphics[width=0.49\columnwidth]{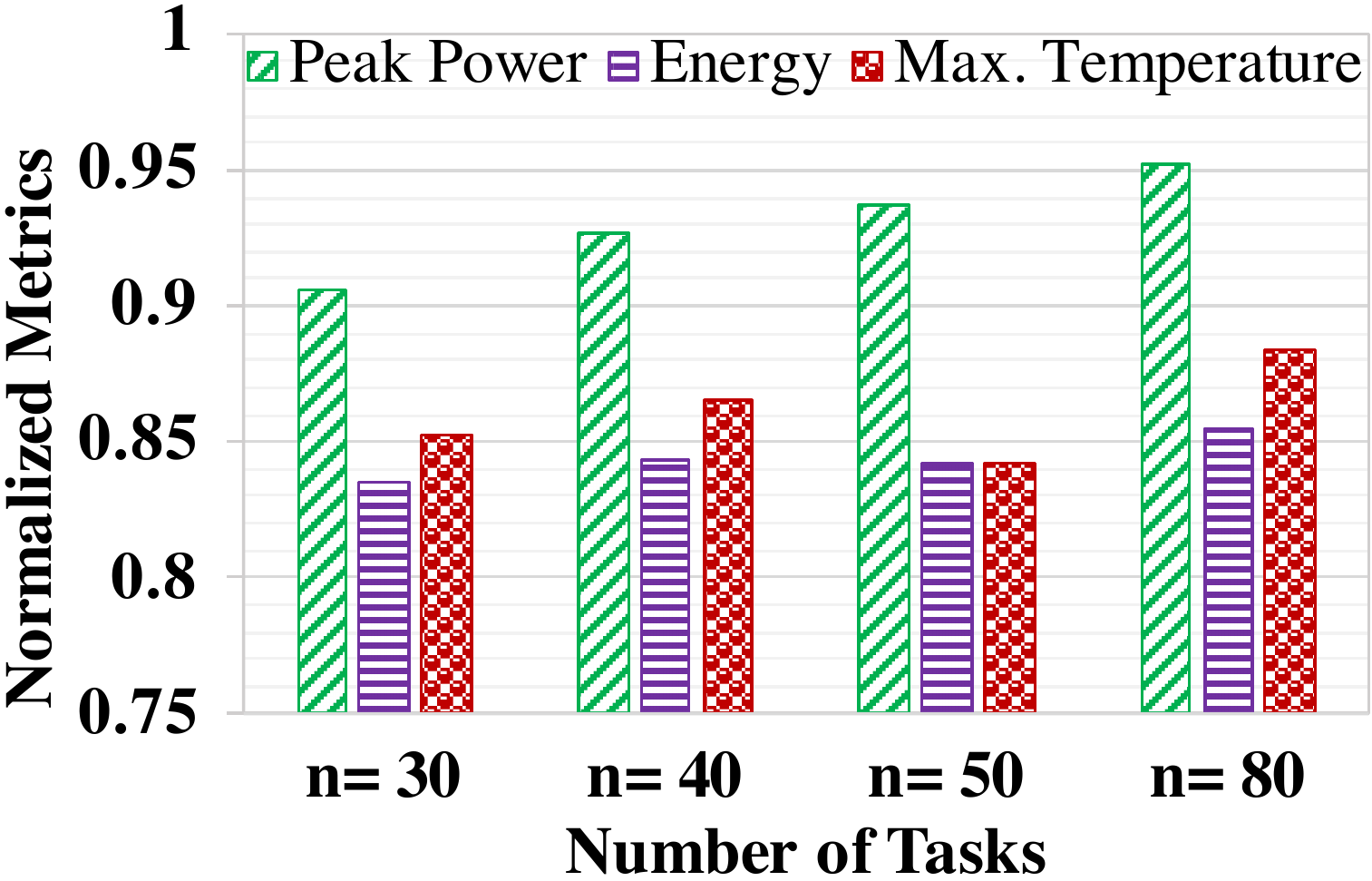}\label{fig:AllTasks}}
        \hfil
        \subfloat[Varying Edge Percentage]{\includegraphics[width=0.47\columnwidth]{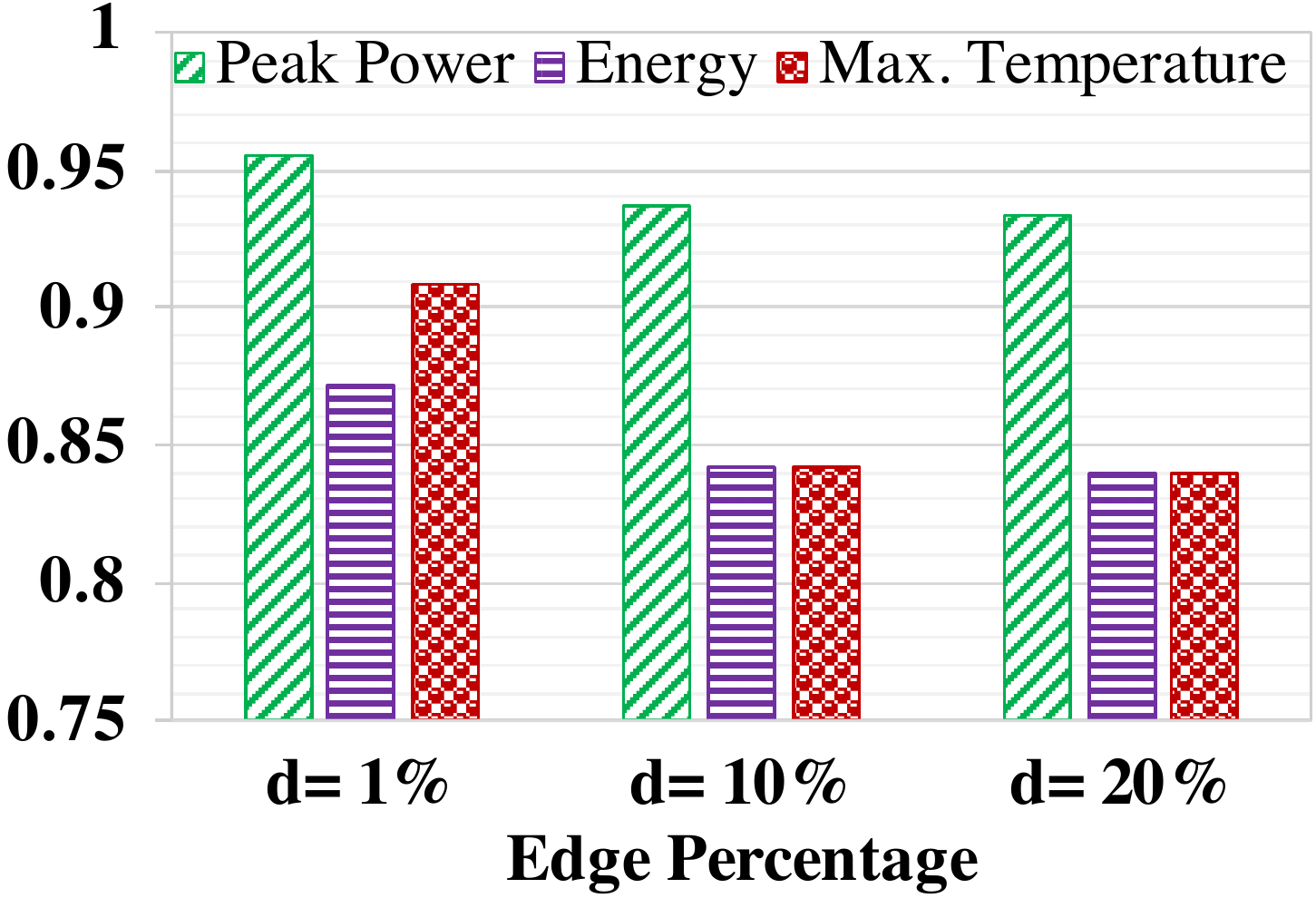}\label{fig:AllEdge}}
		\caption{The improvements in peak power, energy and max. temperature in different scenarios normalized to~\cite{Medina2018}.}
		\label{fig:Scenario}
	\end{minipage}
\end{figure}


\vspace{5pt}

\subsubsection{\textbf{The analysis of the proposed method on peak power, energy and temperature improvement}}

\begin{figure*}[t]
\centering

\subfloat[Varying Utilization Bound]{\includegraphics[width=0.33\linewidth]{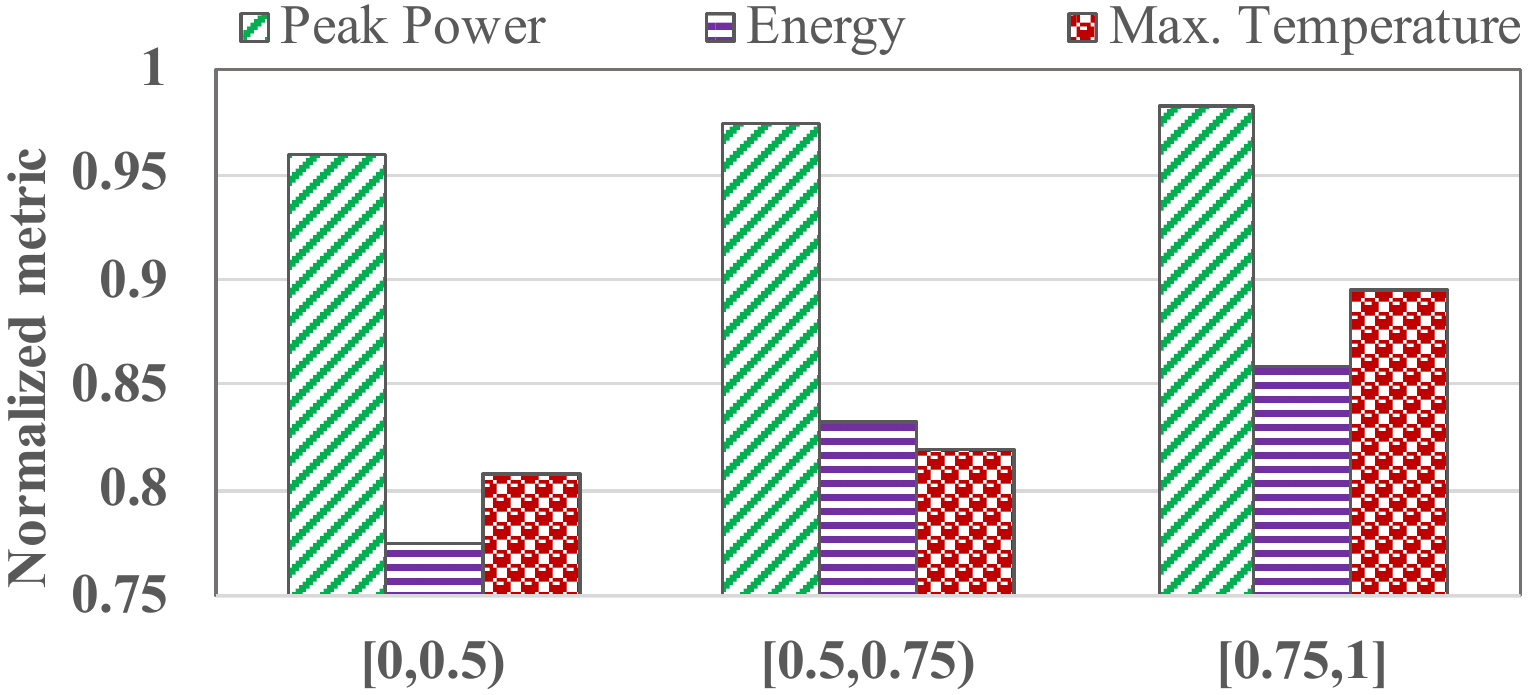}\label{fig:AllCoreshet}}
\hfil
\subfloat[Varying Number of Tasks]{\includegraphics[width=0.32\linewidth]{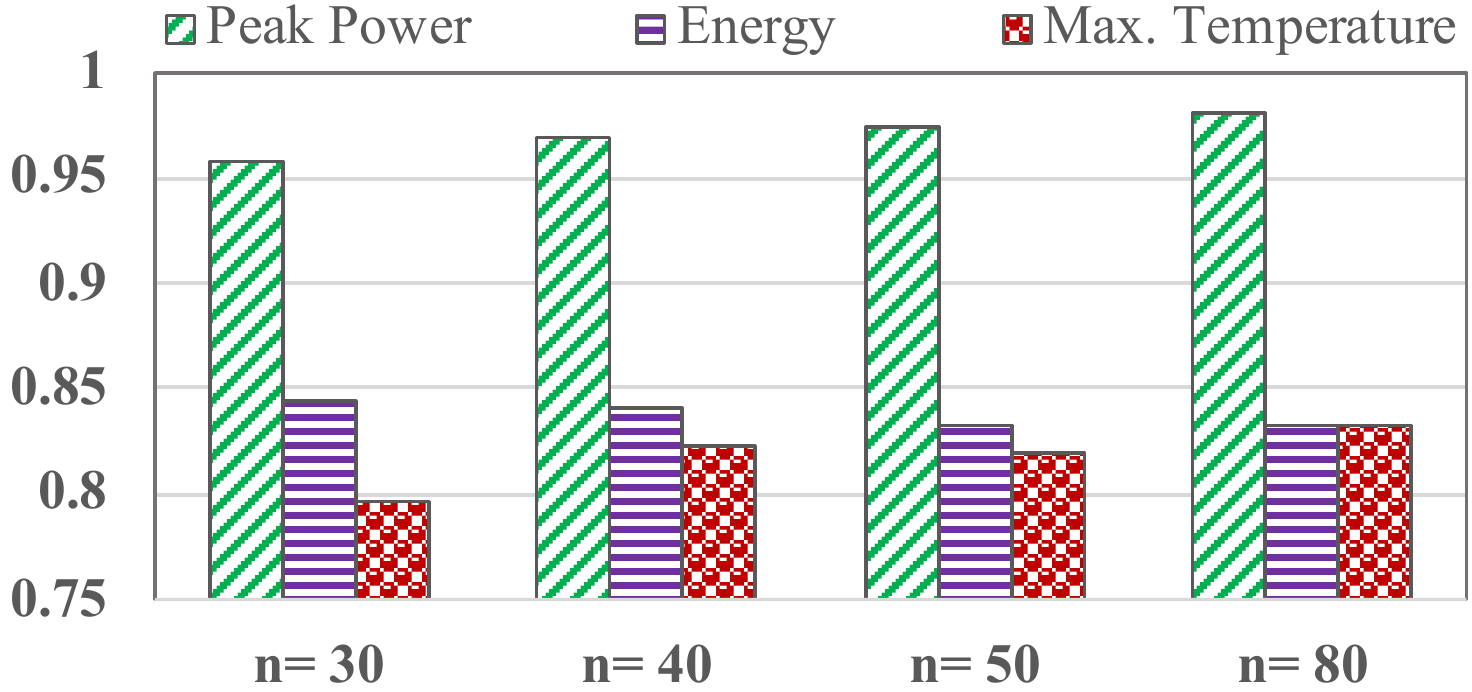}\label{fig:AllUtihet}}
\hfil
\subfloat[Varying Edge Percentage]{\includegraphics[width=0.32\linewidth]{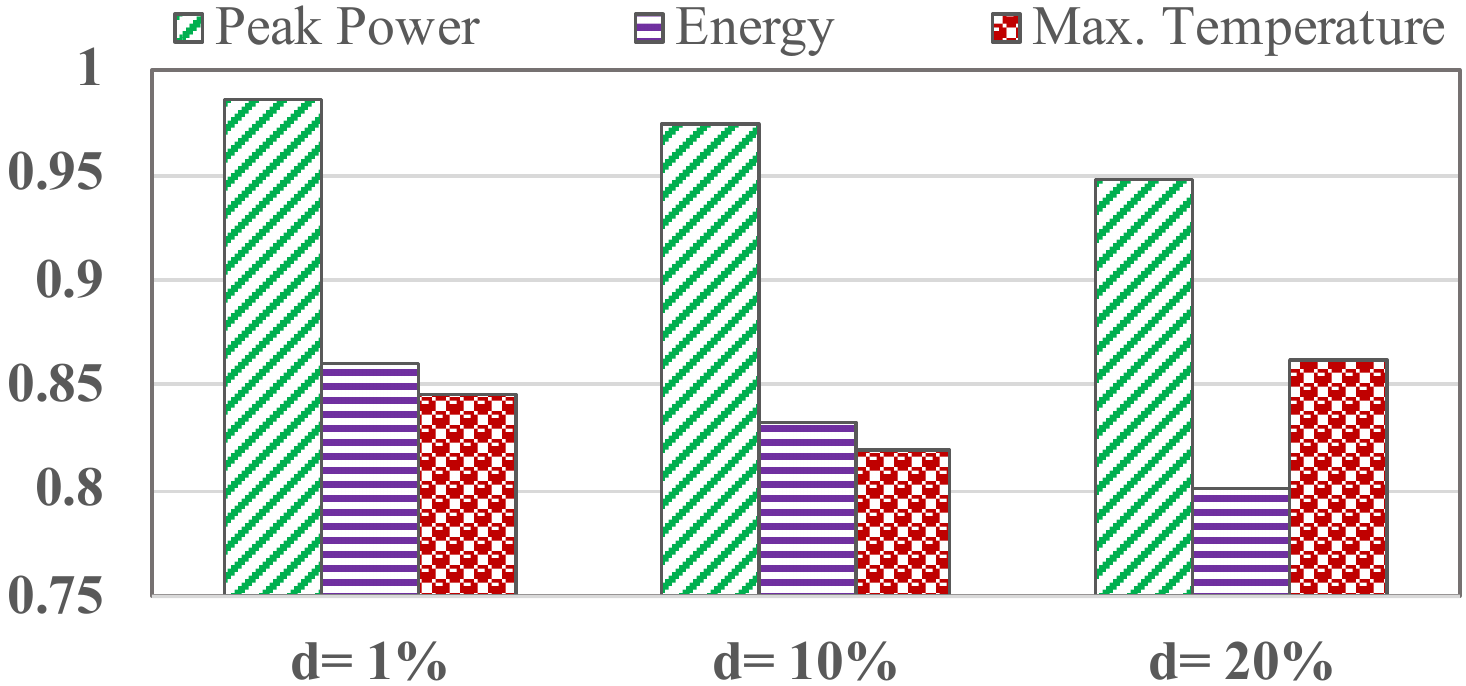}\label{fig:AllTaskshet}}

\caption{Normalized metrics running on the clustered heterogeneous multi-core platforms.
}
\label{fig:clusterresult}
\vspace{-5pt}
\end{figure*}

In order to illustrate how effective our proposed method is with different parameters, we analyze the results under four separate scenarios of Table~\ref{experimenrs}, shown in Fig.~\ref{fig:Scenario}, in which the results are normalized to~\cite{Medina2018}. These results are obtained for multi-core systems, in which there are homogeneous cores based on ARM A7. In general, as the applications get more complicated (e.g., having a large number of tasks or system utilization), it is harder to achieve significant savings in peak power, energy, and maximum temperature. Thanks to our task re-mapping technique where the tasks are redistributed more evenly to the cores at run-time based on their accumulated energy.

For the case of varying the number of cores, since our method only tries to optimize the peak power for each core individually to reduce the time overhead, it is more difficult to maintain a similar system peak power reduction when $c$ is low. Nevertheless, as illustrated in Fig.~\ref{fig:AllCores}, the difference in peak power is significant by increasing the number of cores. In addition, as the temperature of each core is affected by the temperatures of neighboring cores, the reduction in maximum temperature is less by increasing the number of cores. On average, the peak power, maximum temperature, and energy consumption in the system is reduced by $5.015\%$, $17.32\%$, and $15.073\%$, respectively.

The effectiveness of our method depends on the available slacks at run-time and the possibility of assigning them to the tasks. Therefore, if there is less slack due to the nature of the application in terms of the number of tasks and system utilization, the reduction in peak power, energy consumption, and maximum temperature is less. For instance, in Fig.~\ref{fig:AllUti}, when the system utilization is getting higher, the idle time of the core between two consecutive tasks is getting smaller. The tasks also tend to execute longer. Thus, the amount of slacks that can be exploited at run-time is limited. But, overall, the peak power is reduced by at least $3.905\%$, and up to $8.59\%$ in this scenario. Similarly, when there are more tasks in the system with the same $U/c$, the dynamic slacks incurred when the tasks finish earlier than their WCETs are smaller. The reason is that as the expected execution times of the tasks are decreased, the absolute differences between their actual execution time and WCETs are inherently small. However, as seen in Fig.~\ref{fig:AllTasks}, our method manages to reduce the peak power, energy, and maximum temperature on average by $6.96\%$, $15.61\%$ and $13.91\%$, respectively.

Besides, the possibility of releasing the tasks earlier than their presumed start times also affects the outcomes. When the dependency between the tasks is high, a significant amount of them cannot be released earlier. This behavior can either have a positive or negative impact on the system. For the former, the cores might have more idle time because the tasks have to wait longer for their predecessors to finish. For the latter, our method has less opportunity to apply DVFS to tasks. However, at run-time, these idle periods might overlap with the other tasks with the already reduced \textit{V-f} level. The peak power of the whole system is consequently reduced. It can be seen in Fig.~\ref{fig:AllEdge} that, when $d = 20\%$, the best peak power and maximum temperature reduction is achieved compared the cases where $d = 1\%$ and $d = 10\%$.

\vspace{5pt}
    
\subsubsection{\textbf{The analysis of the proposed method on peak power, energy and temperature improvement in a multi-core platform based on the ODROID-XU3 architecture}}

In this section, we analyze the improvement of peak power, energy consumption, and maximum temperature in a clustered heterogeneous multi-core processor in which there are four big (A15) and four LITTLE (A7) cores. 
Here, we have a common \textit{V-f} level for all cores within the same cluster (cluster with big cores or cluster with LITTLE cores), while in the previous results, the frequency of each core was changed individually. We show the results in Fig.~\ref{fig:clusterresult}, in which the improvements in peak power, energy consumption and maximum temperature in the clustered heterogeneous multi-core system across all experiments are up to 5.25\%, 22.44\%, and 20.33\%, respectively in comparison with~\cite{Medina2018}. 
The trends for the clustered heterogeneous multi-core architecture are similar to that obtained for the homogeneous architecture in the previous sub-section. The only notable difference is in the peak power, which is
on average 3.461\% worse than the tasks on a homogeneous multi-core system, due to enforcing of common \textit{V-f} level for the entire cluster. Therefore, the power improvement is somewhat lower.

\begin{figure*}[t]
\centering
\subfloat[Power trace of the big cluster]{\includegraphics[width=0.31\linewidth]{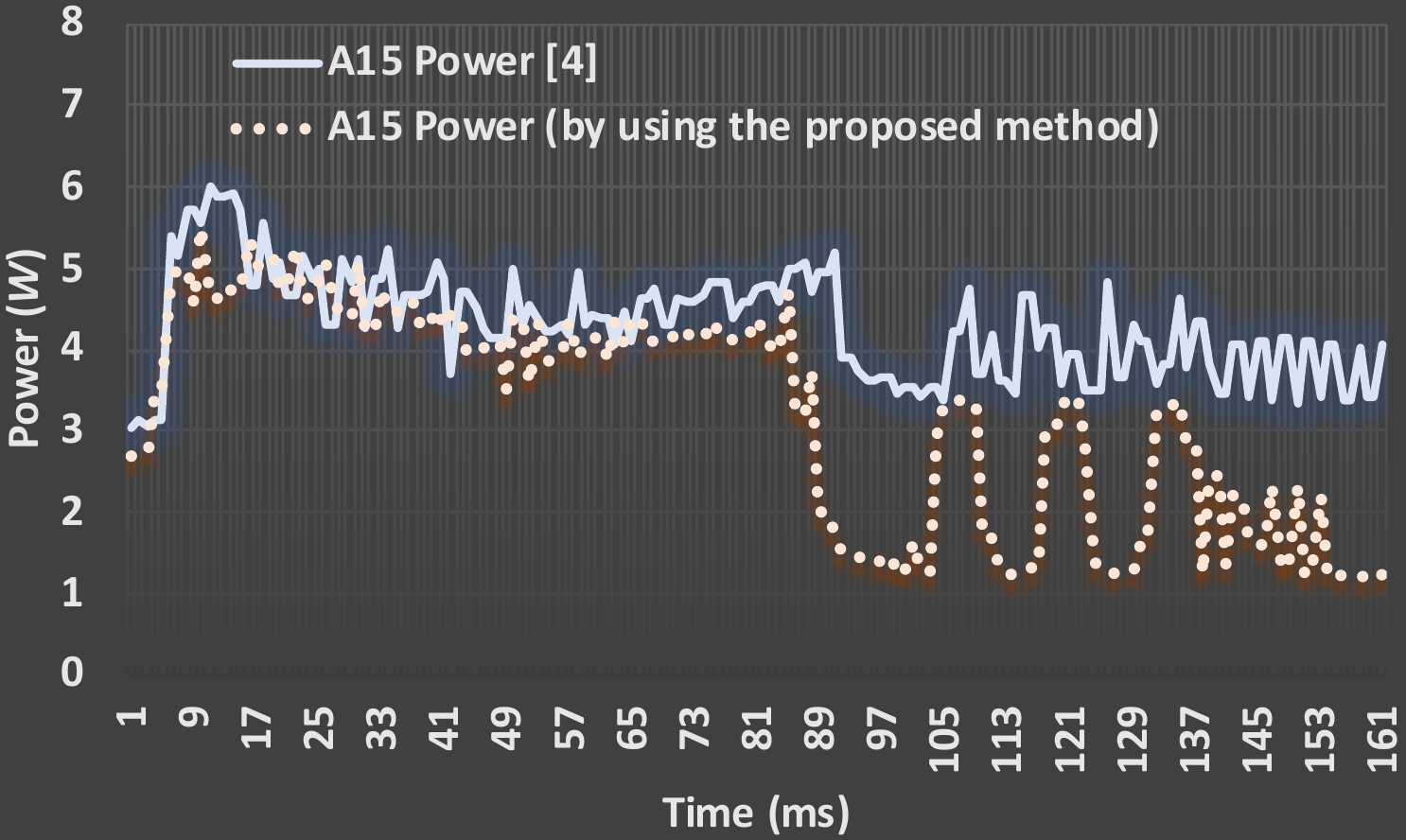}\label{fig:powerA15}}
\hfil
\subfloat[Temperature trace of A15-core2]{\includegraphics[width=0.32\linewidth]{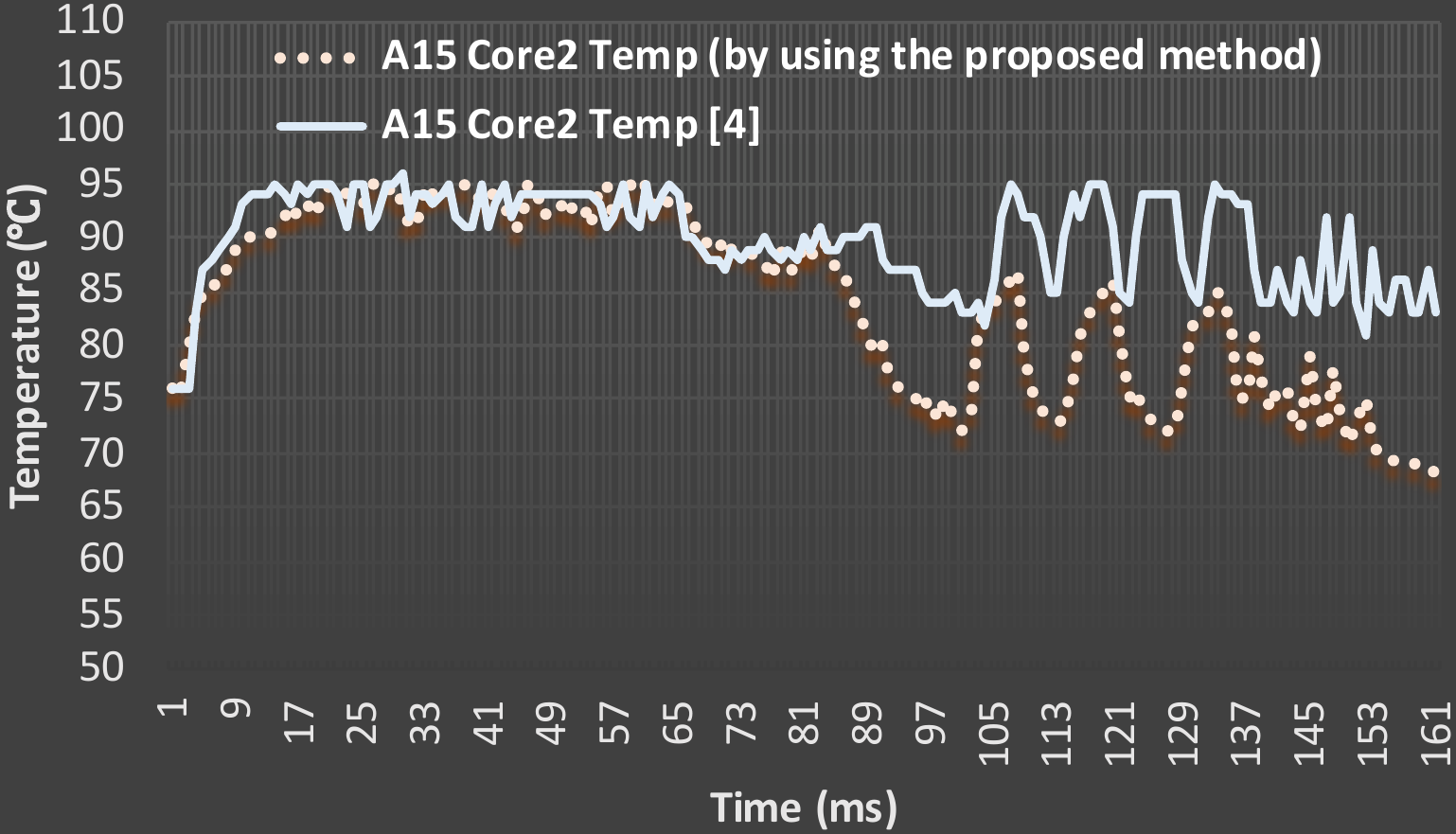}\label{fig:Core1}}
\hfil
\subfloat[Temperature trace of A15-core3]{\includegraphics[width=0.32\linewidth]{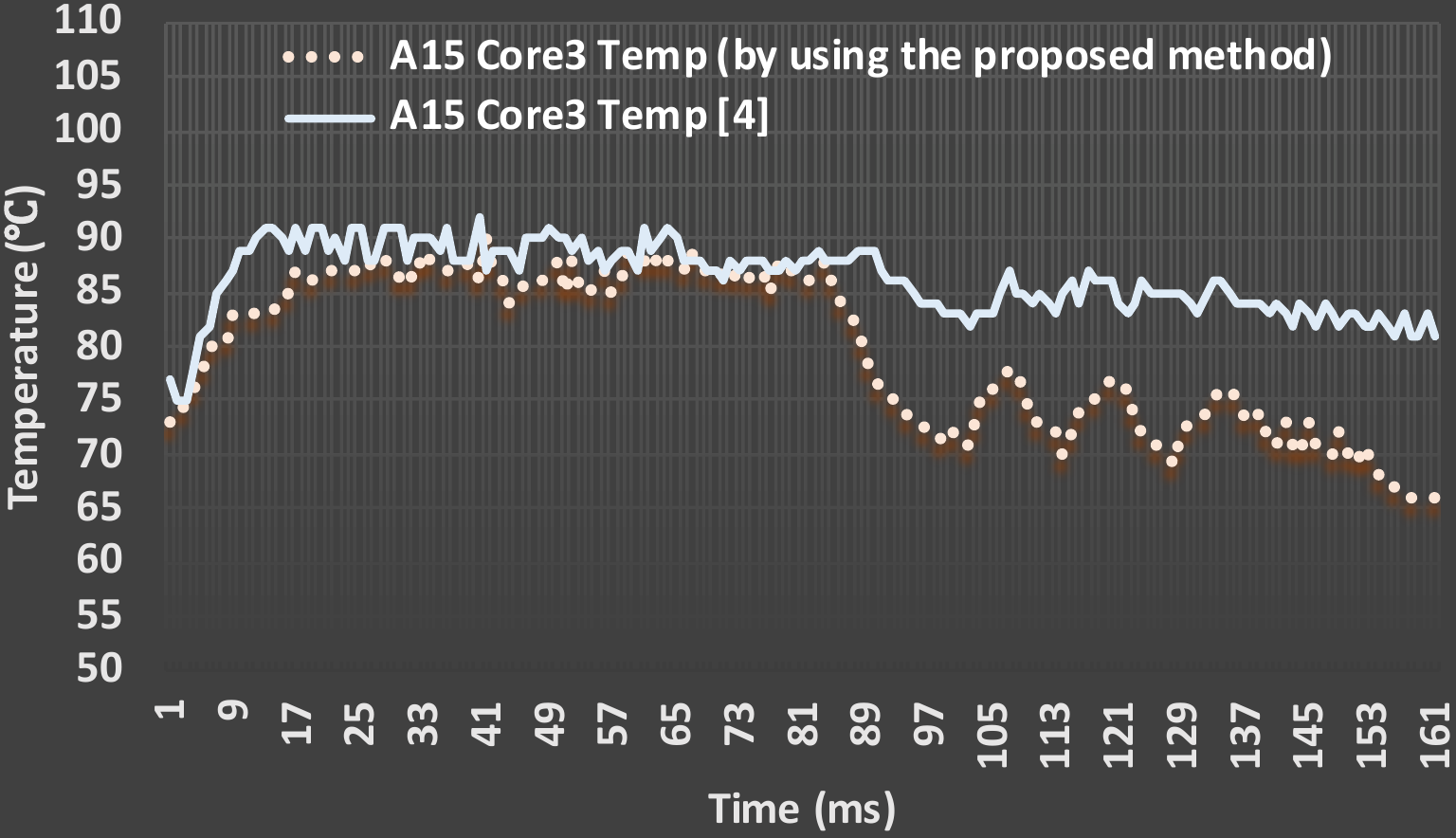}\label{fig:Core2}}

\caption{Power and temperature sensor data of the ODROID~XU3, by running the real MC task graph (Unmanned Air Vehicle).
}
\label{fig:realresult}
\vspace{-5pt}
\end{figure*}

\vspace{5pt}
\subsubsection{\textbf{Evaluation of running real MC task graph (Unmanned Air Vehicle) on real platform}}

Now, we validate the proposed online technique with a real-life application task graph, presented in Fig.~\ref{fig:a_taskGraph}, running on the ODROID~XU3. In particular, we evaluate the impacts of changing frequency on the system power and temperature in this section. The UAV application consists of seven dependent tasks executed on two cores, which has been presented in Fig.~\ref{fig:a_taskGraph}. Since there are no available real benchmarks for the tasks of the graph, we used different benchmarks of Mibench \cite{Mibench} as tasks of the graph. Then, we obtain the WCETs and maximum power consumption of each task with running on the ODROID~XU3. Since WCET analysis is a complicated task~\cite{Bazzaz2020}, we used an existing WCET estimation tool called OTAWA~\cite{OTAWA2010}, to capture the highest WCETs ($C_i^{HI}$). In addition, we run each benchmark, 10,000 times and select the maximum of the measured execution time as the lowest WCET ($C_i^{LO}$). To analyze the system temperature, we run the application on Core2 and Core3 that in general have a higher temperature due to their proximity to the memory and other components. Hence, there is a temperature sensor for each big core and a power sensor for each cluster of ODROID~XU3. Therefore, the power and temperature data of this section are exploited from the board sensors.

Fig.~\ref{fig:realresult} shows the power trace of the cluster of big cores and temperature trace of two cores during runtime in two scenarios of using our DVFS-based proposed method and presented method of \cite{Medina2018}. At runtime, to analyze a task execution time, we select a docker container to run a task and check the time to be aware of the exact start and finish times of the task. Then, the dynamic slack is computed, the slack between task's actual completion time and its WCET. Fig.~\ref{fig:powerA15} shows the power traces of the method of \cite{Medina2018} and our proposed method by considering two tasks looking ahead that the DVFS has been used from almost 90\textit{ms}. In addition, 
as shown in Fig.~\ref{fig:Core1}~and~\ref{fig:Core2}, in general, the average core temperature has been decreased by using the DVFS-based proposed method and looking two tasks ahead. Based on the scheduling of tasks in Fig.~\ref{fig:MExamp}, one of the cores is active until near the middle of the period. However, the temperature of each core is affected by the temperature of neighboring cores. In addition, after executing two tasks in each core and using the dynamic slack to reduce the speed, the cores temperatures are decreasing. Besides, in a part of the task graph period, only Core2 is active but still has high temperatures. Therefore, after applying the proposed method and reducing the \textit{V-f} levels, the cores temperatures are reduced. The proposed method will be more effective and have a significant improvement if there are more tasks that are run on a system with more cores.

\section{Conclusion and Future Work}
\label{Conclusion}
In this article, we studied online peak power and peak temperature reduction in mixed-criticality embedded systems and analyzed the proposed run-time power-aware scheduler on clustered multi-core real platforms. Our presented method uses re-mapping technique and DVFS at runtime whenever there is a dynamic slack. We also proposed the associated cost functions to select the most appropriate task to assign the dynamic slacks to decrease its \textit{V-f} level or to re-map it to another core. We showed that by increasing the number of tasks to look ahead, more peak power and maximum temperature reduction are achieved. In addition, the proposed power-aware scheduler was analyzed in terms of run-time timing overhead in different multi-core platforms. We focused on reducing the run-time scheduler latency to have more usage of dynamic slack and, consequently, more peak power and maximum temperature reduction, while the tasks' deadlines are guaranteed. The results show up to 5.25\%, 20.33\%, and 22.44\% reduction in peak power, peak temperature, and average energy consumption, respectively, compared to recent studies. The proposed solution can manage the system peak power consumption by considering greedily optimizing individual cores. Therefore, as future work, we would consider the management of peak power and thermal cycling issues by considering the whole system to have better improvement in peak power consumption.

\ifCLASSOPTIONcaptionsoff
  \newpage
\fi

\bibliographystyle{IEEEtran}
\bibliography{Ref}

\vspace{-15pt}

\begin{IEEEbiography}[{\includegraphics[width=1in,height=1.25in,clip]{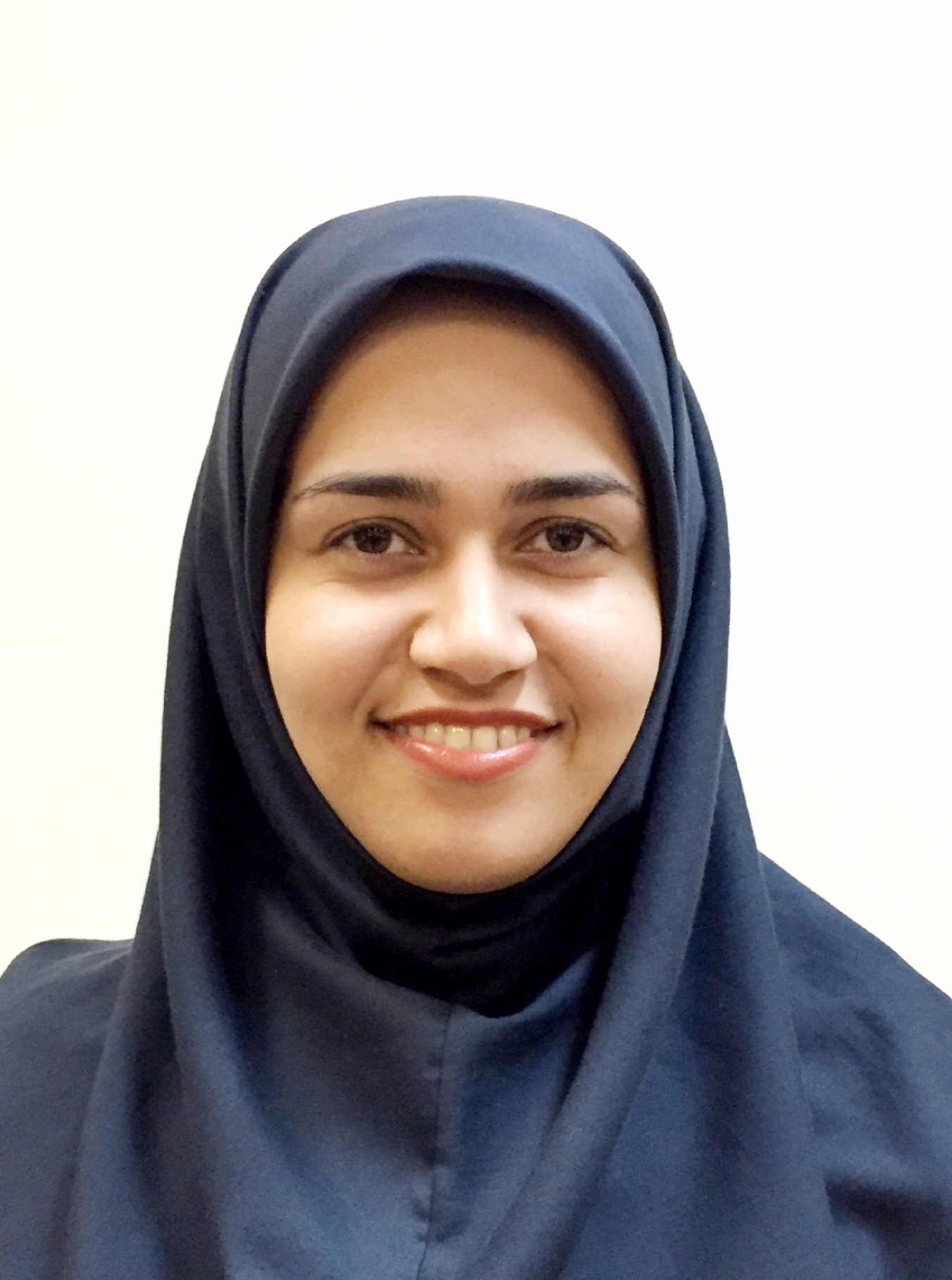}}]{Behnaz Ranjbar} received the B.S. degree in computer engineering from the Amirkabir University of Technology, in 2012, and the M.S. degree from the Sharif University of Technology, Tehran, Iran, in 2014. She is currently pursuing the joint Ph.D. degree with the Sharif University of Technology and the Chair for Processor Design, Technische Universität (TU) Dresden, Dresden, Germany. Her research interest includes real-time systems, fault-tolerant, and low-power embedded systems design.
\end{IEEEbiography}

\vspace{-20pt}

\begin{IEEEbiography}[{\includegraphics[width=0.92in,height=1.25in,clip]{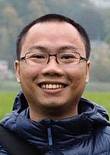}}]{Tuan D. A. Nguyen} has obtained the B.Eng. in computer engineering from Ho Chi Minh City University of Technology, Vietnam, in 2011, and the PhD degree in computer engineering from National University of Singapore, Singapore, in 2018. His interests are in Partially Reconfigurable Heterogeneous Multi-processor System-on-Chip. He was with the Technische Universität Dresden, Dresden, Germany, where he worked as PostDoctoral researcher at Chair for Processor Design from Jan. 2018 to Dec. 2019.
\end{IEEEbiography}

\vspace{-20pt}

\begin{IEEEbiography}[{\includegraphics[width=1in,height=1.25in,clip]{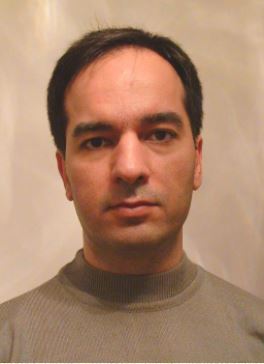}}]{Alireza Ejlali} received the PhD degree in computer engineering from Sharif University of Technology in Tehran, Iran, in 2006. He is currently an associate professor of computer engineering at Sharif University of Technology. From 2005 to 2006, he was a visiting researcher in the Electronic Systems Design Group, University of Southampton, Southampton, United Kingdom. He is currently the Director of the Embedded Systems Research Laboratory, Department of Computer Engineering, Sharif University of Technology. His research interests include low power design, real-time systems, and fault-tolerant embedded systems.
\end{IEEEbiography}

\vspace{-20pt}

\begin{IEEEbiography}[{\includegraphics[width=0.92in,height=1.25in,clip]{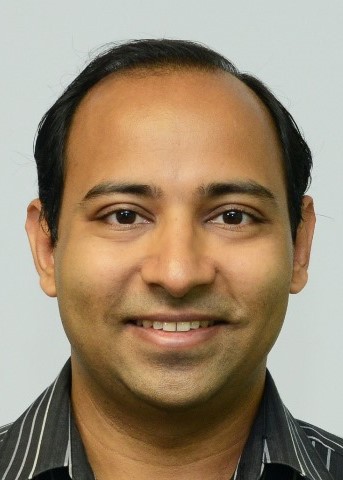}}]{Akash Kumar} (SM’13) received the joint Ph.D. degree in electrical engineering and embedded systems from the Eindhoven University of Technology, Eindhoven, The Netherlands, and the National University of Singapore (NUS), Singapore, in 2009. From 2009 to 2015, he was with NUS. He is currently a Professor with Technische Universität Dresden, Dresden, Germany, where he is directing the Chair for Processor Design. His current research interests include the design, analysis, and resource management of low-power and fault-tolerant embedded multiprocessor systems. 
\end{IEEEbiography}

\clearpage
\appendices
\section{Optimum Solution for Minimizing the Peak Power}

\begin{figure}[t]
\centering
\subfloat[Power trace of the core]{\includegraphics[width=1\columnwidth]{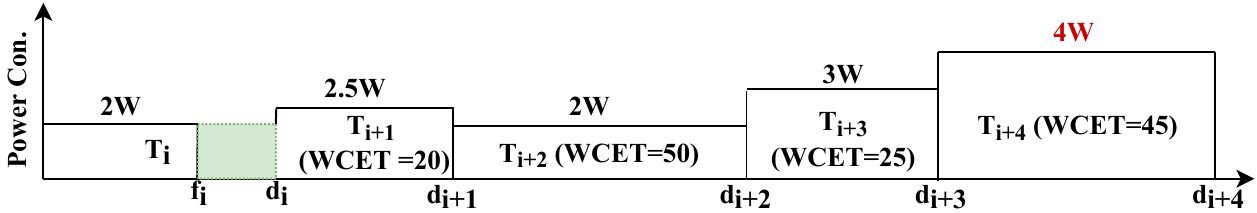}\label{fig:Power}}

\subfloat[Power trace of the core after assigning the slack to the task $T_{i+4}$]{\includegraphics[width=1\columnwidth]{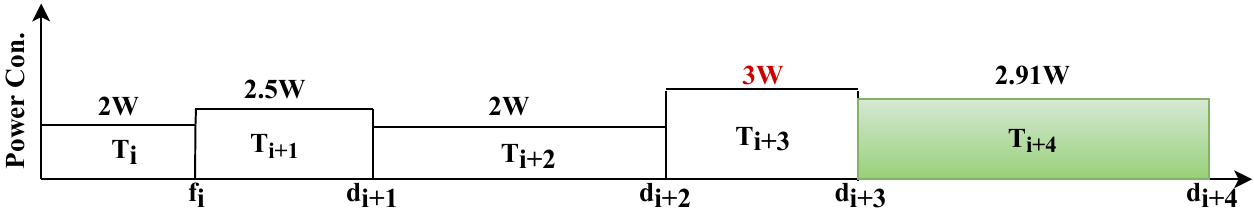}\label{fig:Power1}}

\subfloat[Power trace of the core after assigning the slack to the task $T_{i+1}$]{\includegraphics[width=1\columnwidth]{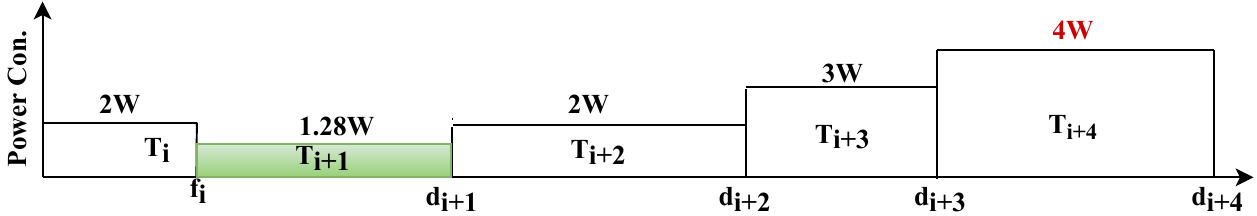}\label{fig:Power2}}
\caption{Power trace of a core}

\label{fig:OptPPow}

\end{figure}

In this paper, we focus on peak power minimization in individual cores during the run-time phase in addition to system peak power reduction. 
Here, we prove that our proposed algorithm is optimal when $\langle\alpha,~\beta\rangle=\langle0,1\rangle$ in Eq.~\ref{eq:4} to select the task solely based on its peak power consumption.
Let us assume that the task $T_i$ finishes its execution at time $f_i$, ahead of its deadline
$d_i$, and a dynamic slack ($S_{dyn}=d_i-f_i$) is generated. The algorithm looks $k$ tasks after generated slack to select the appropriate task and use the generated slack to reduce its \textit{V-f} level and, consequently, decrease its power consumption. 
Without loss of generalization, assume that task $T_{i+l}$ consumes the highest peak power in the core within the $k$ tasks looking ahead, presented in Eq.~\ref{eq:powcore}. This equation can be rewritten as Eq.~\ref{eq:powcore1}, in which $Max(T_{i+1}^{pow},...,T_{i+l-1}^{pow},T_{i+l+1}^{pow},...,T_{i+k}^{pow})<T_{i+l}^{pow}$.
\begin{align}
\label{eq:powcore}
Pow_{core}^{max}|_{[d_i,d_{i+k}]} = Max (T_{i+j}^{pow})|{_{{j=1:k}}} = {T_{i+l}^{pow}},{1\leq l\leq k} 
\end{align}
\begin{align}
\label{eq:powcore1}
&Pow_{core}^{max}|_{[d_i,d_{i+k}]} = \nonumber \\
&Max (Max(T_{i+1}^{pow},...,T_{i+l-1}^{pow},T_{i+l+1}^{pow},...,T_{i+k}^{pow}),T_{i+l}^{pow}) 
\end{align}

If $T^{pow^{'}}_{i+l}$ is the maximum power consumption of task~$T_{i+l}$ after reclaiming the slack and reducing the \textit{V-f} level, then $T^{pow^{'}}_{i+l}<T^{pow}_{i+l}$, therefore, the core's maximum power consumption can be written as follows, which is less than $T^{pow}_{i+l}$:
\begin{align}
\label{eq:powcoren}
&Pow_{core}^{max}|_{[d_i,d_{i+k}]} = \nonumber \\
&Max (Max(T_{i+1}^{pow},...,T_{i+l-1}^{pow},T_{i+l+1}^{pow},...,T_{i+k}^{pow}),T_{i+l}^{pow^{'}}) 
\end{align}

If we select one of the other task between $\{T_{i+1},...,T_{i+l-1},T_{i+l+1},...,T_{i+k}\}$, reduce its \textit{V-f} level and consequently, its power consumption, then the peak power of the core is still limited by $T^{pow}_{i+l}$ according to Eq.~\ref{eq:powcore1}. Hence, this power consumption is more than the optimum power consumption obtained by Eq.~\ref{eq:powcoren}. 
$T_{i+l}$ is, therefore, the optimum task to which the slack should be assigned (given the constraint that all the slack is assigned to \textit{one} of the following $k$ tasks). We conclude that whenever a dynamic slack is generated, the proposed approach for selecting the appropriate task provides the optimum solution to minimize the peak power consumption of individual cores in the run-time phase.

Fig.~\ref{fig:OptPPow} shows a part of a static schedule of tasks on a core. Based on the peak power consumption of tasks in Fig.~\ref{fig:Power}, task $T_{i+4}$ is the appropriate task which consumes the highest peak power in the core in the time interval [$d_i,d_{i+4}$]. Therefore, assigning the slack to this task will lower the peak power to below 4W (If we have a dynamic slack ($S_{d}=d_i-f_i=5$), then $Power_{core}^{max}=3W$ after slack assignment, shown in Fig.~\ref{fig:Power1}). If we assign the generated slack to one of the other tasks (e.g., $T_{i+1}$) instead, then the peak power of the core is still limited by $T_{i+4}$, i.e. 4W, as can be seen in Fig.~\ref{fig:Power2}.

\section{Finding the Optimum Number of Look-Ahead Tasks by Varying the Tasks' Properties}

Here, we show the relation between the number of look-ahead tasks ($k$) and the task property, edge percentage ($d\%$) to model the system capability such as peak power minimization, energy consumption, and maximum temperature. For this analysis, the data from the experiments with four cores ($c$), and the system utilization per core ($U/c$) in the range of [0.5,0.75) is used. Average data of 100 task set runs has been used.

\begin{figure}[t]
\centering
\includegraphics[width=0.9\columnwidth]{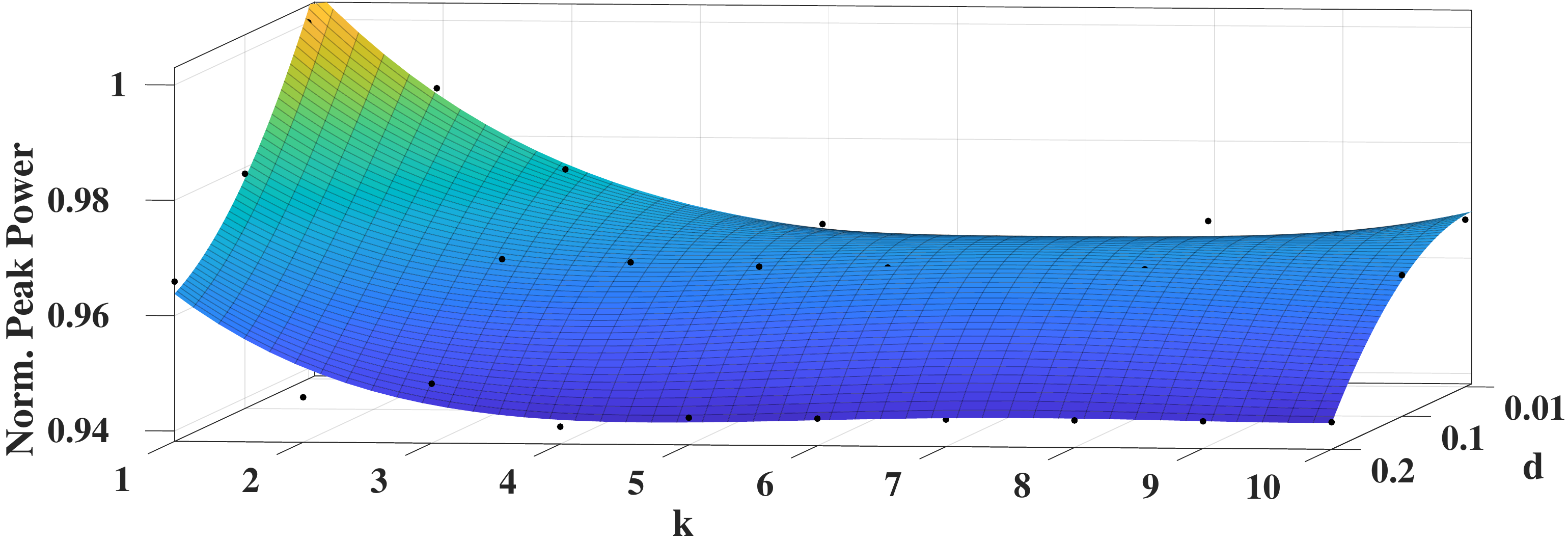}
\caption{Impact of number of look-ahead tasks and edge percentage on normalized peak power consumption.}
\label{fig:equation3D}
\vspace{-5pt}
\end{figure}

We use the Matlab Curve Fitting Tool to derive the polynomial functions of various system parameters. Fig.~\ref{fig:equation3D} shows the curve of the system peak power consumption by varying $k$ and $d$ normalized to the result for $k=1$ and the corresponding equation is shown in Eq.~\ref{eq:findk}. 
This equation is the polynomial function with the maximum degree of four with the minimum Root Mean Square Error (RMSE), equal to 0.0024. 
\begin{align}
\label{eq:findk}
Pow_{peak}^{Norm.} (k,d) = 1.033 -0.5082d - 0.03374k + 1.332d^2 + \nonumber \\
0.1799dk + 0.006184k^2 - 0.7267d^2k - 0.01158dk^2 -  \nonumber \\
0.0005375k^3 + 0.05294d^2k^2 - 0.0001314dk^3 + 1.912\times10^{-5}k^4
\end{align}

\begin{figure}[ht]
\centering
\includegraphics[width=0.7\columnwidth]{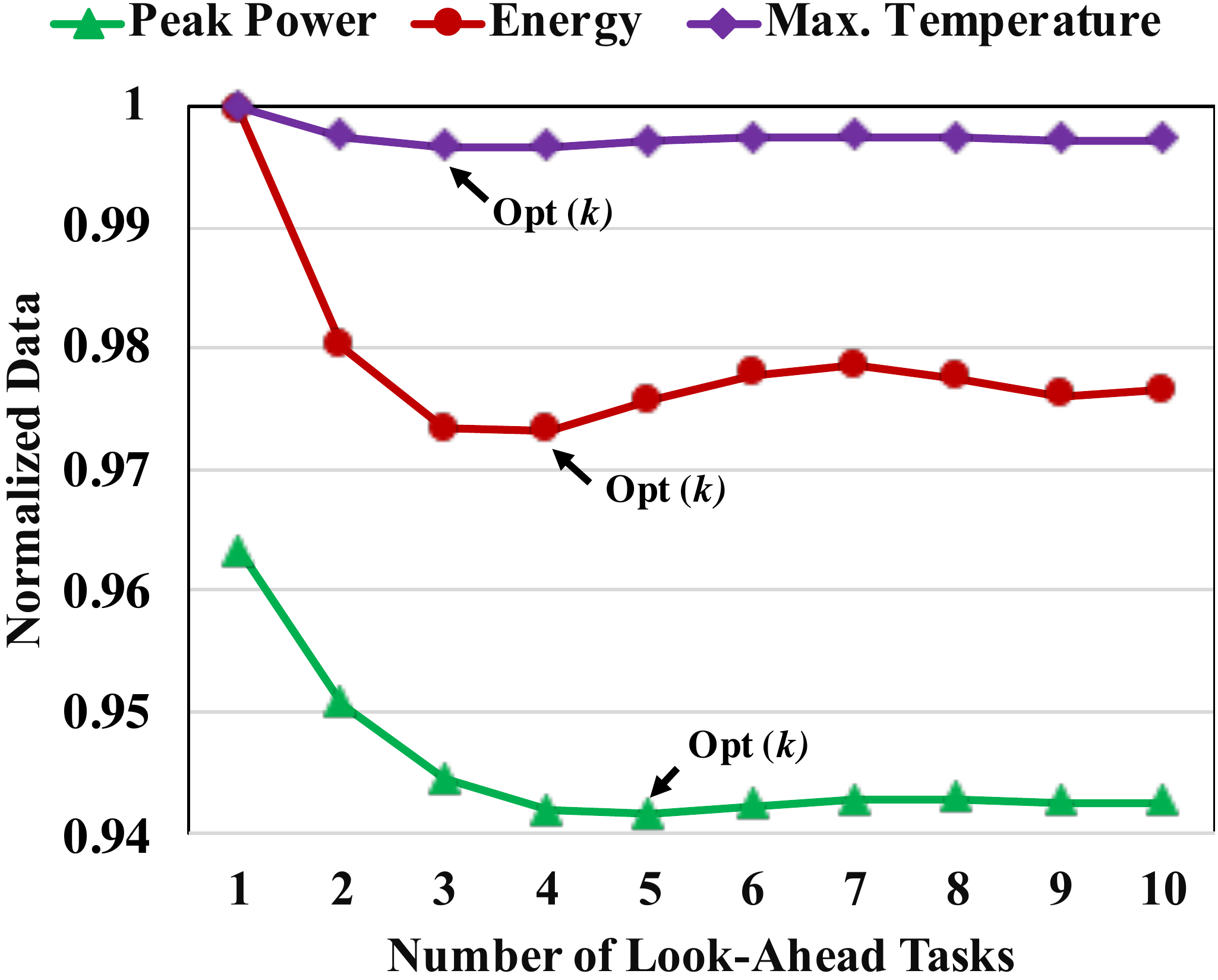}
\caption{Impact of number of look-ahead tasks on normalized peak power, energy and max. temperature while $d=20\%$.}
\label{fig:Optk20percent}
\end{figure}

The equation above can also be used to mathematically derive the optimal $k$ for a particular task property to optimize the various metrics. For example, if $d$ is kept as 20\% in Eq.~\ref{eq:findk}, the minimum value of the curve is obtained when $k=5$ which is shown in Fig.~\ref{fig:Optk20percent}. In addition, by deriving the corresponding equations for the normalized maximum temperature (Eq.\ref{eq:findkE}) and energy consumption (Eq.\ref{eq:findkTemp}), the optimum value of $k$ for the system's max. temperature is $k=3$ and for the energy consumption is $k=4$, as shown in Fig.~\ref{fig:Optk20percent}.
\begin{align}
\label{eq:findkE}
Energy_{peak}^{Norm.} (k,d) = 0.9347 + 1.176d - 0.05964k - 3.304d^2 \nonumber \\
- 0.01962dk + 0.01355k^2 + 0.3278d^2k + 0.005128dk^2 - \nonumber \\
0.001431k^3 - 0.02466d^2 k^2 - 0.000231dk^3 + 5.616\times 10^{-5}k^4
\end{align}
\begin{align}
\label{eq:findkTemp}
T_{peak}^{Norm.} (k,d) = 0.9925 + 0.07826d - 0.006975k - 0.08123d^2 \nonumber \\ 
- 0.008346dk + 0.001644k^2 + 0.05333d^2k + 0.001238dk^2 \nonumber \\
- 0.0001762k^3 - 0.003814d^2k^2 - 10^{-5}(4.653dk^3 + 0.6901k^4)
\end{align}

\newpage

\end{document}